\documentclass{emulateapj}
\usepackage[dvips]{color}
\usepackage{multirow}
\usepackage{amsmath}
\usepackage{subfloat}
\newcommand{\Ha}{H$\alpha$}

\newcommand{\OIII}{[O\,{\sc iii}]\,$\lambda$}

\newcommand{\OI}{[O\,{\sc i}]}

\newcommand{\NII}{[N\,{\sc ii}]}
\newcommand{\CII}{[C\,{\sc ii}]}

\newcommand{\LOI}{$L_{\rm [O\,{\scriptsize \textsc{i}}]63\,\mu{\rm m}}$}
\newcommand{\LOIFIR}{$L_{\rm [O\,{\scriptsize \textsc{i}}]63\,\mu{\rm m}}$}
\newcommand{\OIFIR}{[O\,{\sc i}]\,63\,\mum}

\newcommand{\LCII}{$L_{\rm [C\,{\scriptsize \textsc{ii}}]}$}
\newcommand{\LIR}{$L_{\rm IR}$}
\newcommand{\LFIR}{$L_{\rm FIR}$}

\newcommand{\mum}{$\mu$m}
\newcommand{\LOIII}{$L_{\rm O\,[{\scriptsize \textsc{iii}}]}$}
\newcommand{\NIIIO}{[N\,{\sc ii}]205\,$\mu$m}

\newcommand{\fircolor}{$f_{60}/f_{100}$}
\newcommand{\herschel}{{\it Herschel}}
\newcommand{\wise}{{\it WISE}}
\newcommand{\iras}{{\it IRAS}}
\newcommand{\lsolar}{$L_\odot$}

\newcommand{\OIIIDOUBLE}{[O\,{\sc iii}]$\lambda\lambda$4959,\,5007}
\newcommand{\OIDOUBLE}{[O\,{\sc i}]$\lambda\lambda$6300,\,6364\,\AA}
\newcommand{\SIIDOUBLE}{[S\,{\sc ii}]$\lambda\lambda$6716,\,6731}
\newcommand{\NIIDOUBLE}{[N\,{\sc ii}]$\lambda\lambda$6548,\,6583}
\newcommand{\kms}{km s$^{-1}$}

\begin{document}
\shorttitle{Properties of ISM in IR Bright QSOs}
\shortauthors{Zhao et al.}

\title{Properties of Interstellar Medium In Infrared Bright QSOs Probed by \OI\,63\,$\mu${\MakeLowercase m} and \CII\,158\,$\mu${\MakeLowercase m} Emission Lines$^\star$}
\author{Yinghe Zhao\altaffilmark{1, 2, 3}, Lin Yan\altaffilmark{1} and Chao-Wei Tsai\altaffilmark{4,5}}
\altaffiltext{1}{Infrared Processing and Analysis Center, California Institute of Technology 100-22, Pasadena, CA 91125, USA; zhaoyinghe@gmail.com}
\altaffiltext{2}{Purple Mountain Observatory, Chinese Academy of Sciences, Nanjing 210008, China}
\altaffiltext{3}{Key Laboratory of Radio Astronomy, Chinese Academy of Sciences, Nanjing 210008, China}
\altaffiltext{4}{Jet Propulsion Laboratory,  California Institute of Technology, 4800 Oak Grove Dr., Pasadena, CA 91109, USA}
\altaffiltext{5}{NASA   Postdoctoral   Program   Fellow}
\altaffiltext{$\star$}{\herschel\ is an ESA space observatory with science instruments provided by European-led Principal Investigator consortia and with important participation from NASA.}
\date{Received:~ Accepted:~}
\begin{abstract}
We present a study of interstellar medium in the host galaxies of 9 QSOs at 0.1\,$<$\,$z$\,$<$\,0.2 with blackhole masses of $3\times10^7\,M_\odot$ to $3\times10^9\,M_\odot$ based on the far-IR spectroscopy taken with {\it Herschel Space Observatory}. We detect the \OI\,63$\mu$m (\CII\,158$\mu$m) emission in 6(8) out of 8(9) sources. Our QSO sample has far-infrared luminosities (\LFIR) $\sim$\,several times $10^{11}L_\odot$. The observed line-to-\LFIR\ ratios (\LOI/\LFIR\ and \LCII/\LFIR) are in the ranges of 2.6\,$\times$\,$10^{-4}$\,$-$\,$10^{-2}$ and 2.8\,$\times$\,$10^{-4}$\,$-$\,2\,$\times$\,$10^{-3}$ respectively (including upper limits). These ratios are comparable to the values found in local ULIRGs, but higher than the average value published so far for $z$$>$1 IR bright QSOs.  One target, W0752+19, shows an additional broad velocity component ($\sim$720\,km/s), and exceptionally strong \OI\,63$\mu$m emission with \LOI/\LFIR\ of $10^{-2}$, an order of magnitude higher than that of average value found among local (U)LIRGs. Combining with the analyses of the {\it SDSS} optical spectra, we conclude that the \OI\,63$\mu$m emission in these QSOs is unlikely excited by shocks. We infer that the broad \OIFIR\ emission in W0752+19 could arise from the warm and dense ISM in the narrow line region of the central AGN. Another possible explanation is the existence of a dense gas outflow with $n_{\rm H}\sim10^4$\,cm$^{-3}$, where the corresponding broad \CII\ emission is suppressed. Based on the far-IR \OI\ and \CII\ line ratios, we estimate the constraints on the ISM density and UV radiation field intensity of $n_{\rm H} \lesssim 10^{3.3}$ cm$^{-3}$ and $10^3<G_0 \lesssim 10^{4.2}$, respectively. These values are consistent with those found in local Seyfert 1 ULIRGs. In contrast, the gas with broad velocity width in W0752+19 has $n_{\rm H} \gtrsim 10^{4.3}$ cm$^{-3}$ and $G_0>10^4$.
\end{abstract}

\section{Introduction}
Galaxy formation and evolution is essentially a complex process of interplay between stars, interstellar medium (ISM) and central black holes. The formation of stars and growth of black holes all require input of ISM, either in the form of cold molecular or multi-phase gas medium.  Conversely, feedback from massive young stars and black hole accretion can inject tremendous energy and momentum back into the ISM, influencing the next rounds of star formation and black hole growth. Therefore, a comprehensive picture of galaxy formation requires good understanding of stars, ISM and central black holes.

ISM in a galaxy can have several different phases -- cold molecular gas, warm neutral gas and ionized gas.  Warm, neutral gas medium is an important part of galaxy ISM. It can be effectively probed by ionic and atomic forbidden fine-structure emission lines in the far-infrared, transitions such as [C\,{\sc ii}]158\,\mum, [O\,{\sc i}]63\,\mum\ and 145\,\mum\ lines.  These lines are important coolant for warm, neutral ISM. Furthermore, [C\,{\sc ii}]158\,\mum\ is the brightest far-infrared line in most galaxies and accounts for for 0.1-1\%\ of the total FIR luminosity (e.g. Stacey et al. 1991; Malhotra et al. 2001; D\'{i}az-Santos et al. 2013; Sargsyan et al. 2014).  The \OIFIR\ emission can be brighter than the \CII\ emission when FIR color, $f_{60}/f_{100}$ (i.e. the 60-to-100 \mum\ ratio), larger than $\sim$0.7 (e.g. Malhotra et al. 2001; Brauher et al. 2008). Finally, far-IR fine-structure line ratios between [C\,{\sc ii}]158\,\mum, [O\,{\sc i}]63\,\mum\ and 145\,\mum\ lines can also provide critical diagnostic tools for inferring physical conditions of ISM, such as temperature, density, and intensity of the radiation field, by comparing observational results with model predictions of photodissociation regions (PDRs; e.g., Tielens \& Hollenbach 1985; Kaufman et al. 1999; Fischer et al. 2014). 

For many years, such detailed studies of neutral ISM using far-IR fine-structure line ratios have been only possible for the Galactic star forming regions and near-by galaxies in the local group.  The observations from {\it Infrared Space Observatory (ISO)} (Kessler et al. 1996, 2003) has revealed for the first time that \CII-to-FIR ratios of nearby normal galaxies are an order of magnitude higher than that of local Ultra-Luminous Infrared Galaxies (ULIRGs; $L_{\rm IR}$\,$>$\,$10^{12}L_\odot$) (Malhotra et al. 2001; Luhman et al. 2003; Brauher et al. 2008).  The exact physical explanations for this difference are still the topic for debate (Malhotra et al. 2001; Abel et al. 2009; D\'{i}az-Santos et al. 2013).

The advent of Atacama Large Millimeter/submillimeter Array (ALMA) has made \CII\ the most utilized emission line for studying high-$z$ galaxies at millimeter wavelengths. The available ALMA frequencies and its extremely high sensitivity have produced an explosion in the number of \CII\ line detections among various types of objects at high redshifts, including star forming galaxies, AGNs, submillimeter selected galaxies (SMG), and QSOs (e.g., Swinbank et al. 2012; Wang et al. 2013; Willott et al. 2013; Decarli et al. 2014; Capak et al. 2015). However, it is challenging to detect multiple FIR fine-structure lines of low redshift objects from the ground because the limited atmospheric windows.

{\it Herschel Space Observatory (Herschel)} (Pilbratt et al. 2010) provides the perfect spectral coverage, resolution and sensitivities for observations of FIR fine-structure lines from galaxies at low redshifts, which are too distant/faint for {\it ISO}  and not accessible from ALMA. {\it Herschel} has already assembled a large FIR spectroscopic dataset for local (U)LIRGs (D\'{i}az-Santos et al. 2013; Farrah et al. 2013; Zhao et al. 2013; Lu et al. 2014; Sargsyan et al. 2014).  Similar studies for $z$\,$>$\,4 QSOs have been done using IRAM telescopes and ALMA (Maiolino et al. 2005; Gallerani et al. 2012; Venemans et al. 2012; Wagg et al. 2012; Leipski et al. 2013; Wang et al. 2013; Willott et al. 2013). However, there are lack of similar observations for infrared bright QSOs at low redshifts.

Therefore, it is important to increase the number of low-$z$ QSOs with FIR fine-structure line observations of low-$z$ QSOs. Even with a small increase in the number, such a study can provide insight into how ISM of QSOs/galaxies evolves from low to high luminosity, and from low-$z$ to high-$z$. Here we report our results of the \CII158\,\mum\ and \OIFIR\ line observations for 9 IR bright QSOs at $z$\,$\sim$\,0.1\,-\,0.2, taken with the Photodetector Array Camera and Spectrometer (PACS; Poglitsch et al. 2010) onboard \herschel.  The sample is selected using both {\it Sloan Digital Sky Survey (SDSS)} and {\it Wide Infrared Survey Explorer (WISE)} (Wright et al. 2010). 

In this paper, we present our PACS spectra of these two lines, extract line and continuum fluxes, and fit the infrared spectral energy distributions (SEDs) for these QSOs. We also compare our line-to-FIR continuum ratios with other types of galaxies and PDR models. The paper is organized as follows: we describe our sample selection, observations and data reduction in Section 2, present our results and discussion in Section 3, and summarize briefly in the last section. Throughout the paper, we adopt a Hubble constant of $H_0=71~$km~s$^{-1}$~Mpc$^{-1}$, $\Omega_{\rm M} =0.27$, and $\Omega_\Lambda=0.73$ (Spergel et al. 2007).
 
\section{Sample, Observations and Data Reduction}
\subsection{The IR Bright QSO Sample}
The IR bright QSOs studied in this paper were identified by using the optical spectra from {\it SDSS} and drawn from the overlapping sky region between {\it WISE} and SDSS DR7 data.  The {\it SDSS} QSO catalog from the 7th data release (Schneider et al. 2010) was used as the initial input. The IR bright QSOs are selected according to: 1). Their {\it SDSS} redshifts are in the range of 0.1 and 0.2; 2) The observed \wise\ $W4$ flux (i.e. at 22\,\mum; hereafter $f_{W4}$), $f_{W4}$\,$>$\,100\,mJy. The first criterion with the narrow redshift slice was chosen so that the major cooling line \CII158\,\mum\ is still well within the sensitive part of the PACS spectral coverage. The second criterion ensures that the selected sources are bright enough for FIR spectroscopy (their 100\,\mum\ fluxes $f_{100}$\,$\geq$\,300\,mJy based on local QSO templates).  There are only 9 sources satisfies these criteria, and the details of the full sample are listed in Table \ref{table:obslog}.

\subsection{Herschel/PACS FIR Spectroscopy}
\setcounter{figure}{0}
\begin{figure*}[pthb]
\centering
\includegraphics[width=0.95\textwidth,bb=80 117 550 735]{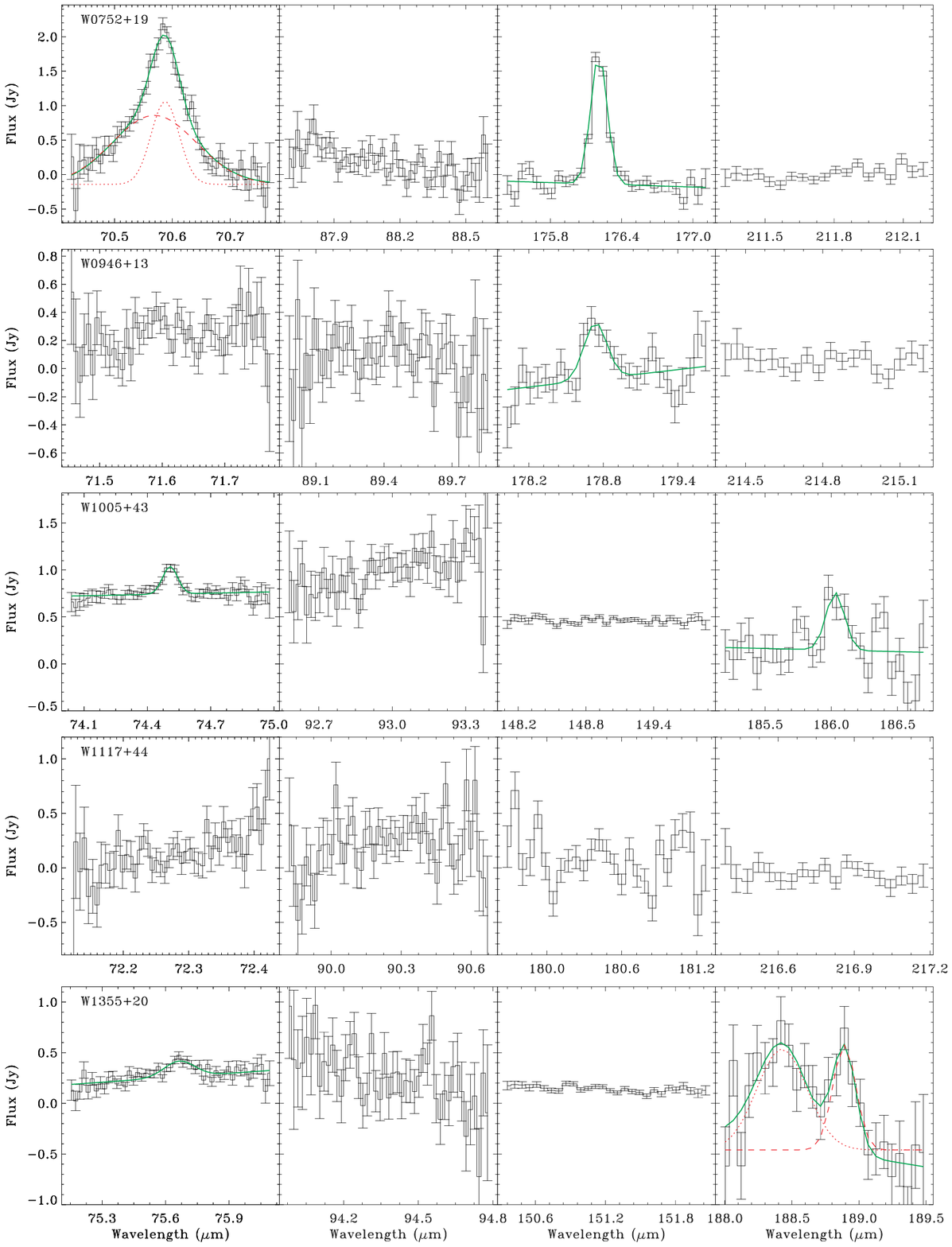}
\caption{PACS spectra for our QSOs in the observed frame (black). The panels with the range of x-axis within $70-76$ and $175-190\,\mu{\rm m}$ show the the \OI\ and \CII\ lines, respectively, and the other two panels display the continuum. For the line emission, we suppose the fitted (pseudo-)continuum and emission line (thick green) if a line was detected, and individual components (shifted arbitrarily in y-axis) are shown with (red) dashed and dotted lines, respectively, if two components fitting for a line applied.}
\label{Fig1}
\end{figure*}

\setcounter{figure}{0}
\begin{figure*}[pthb]
\centering
\includegraphics[width=0.95\textwidth,bb=80 187 550 679]{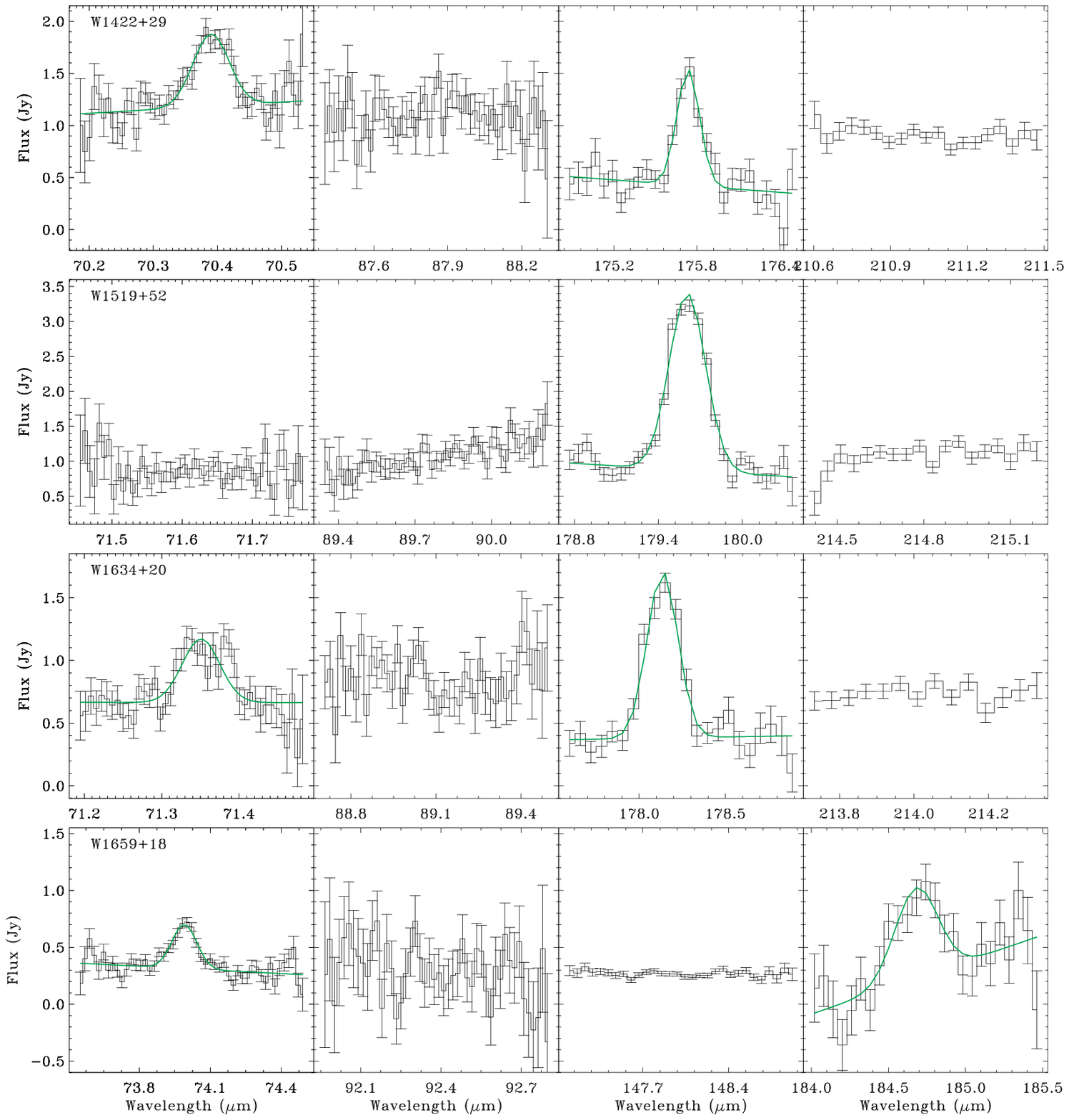}
\caption{Continued}
\label{Fig1cont}
\end{figure*}

\begin{deluxetable*}{lccccclll}[pthb]
\tablecaption{The Sample and PACS Observation Log\label{table:obslog}}
\tablewidth{0pt}
\tablehead{
\colhead{Galaxy}&\colhead{R.A. (J2000)}&\colhead{Decl. (J2000)}&\colhead{Redshift}&\colhead{line{\tablenotemark{a}}}&\colhead{Exposure (s)\tablenotemark{b}}&\colhead{ObsID}&\colhead{Obs Date}\\
\colhead{(1)}&\colhead{(2)}&\colhead{(3)}&\colhead{(4)}&\colhead{(5)}&\colhead{(6)}&\colhead{(7)}&\colhead{(8)}}
\startdata
W0752+19 & 07 52 17.84 & +19 35 42.3  &0.1170 & \OI/\CII &$384\times10\times2$/$344\times6$ & 1342251203/35 & 2012-09-19/20\\
W0946+13 & 09 46 52.58 & +13 19 53.9  &0.1332 & \OI/\CII &$384\times10\times2$/$344\times4$ & 1342256250/49 & 2012-11-29\\
W1005+43 & 10 05 41.87 & +43 32 40.4  &0.1788 & \OI/\CII &$368\times10\times2$/$344\times5$& 1342254963/64 & 2012-11-12\\
W1117+44 & 11 17 06.40 & +44 13 33.3  &0.1437 & \OI/\CII &$384\times10\times2$/$344\times3$& 1342256771/72 & 2012-12-03\\
W1355+20 & 13 55 50.20 & +20 46 14.5  &0.1965 & \OI/\CII &$368\times10\times2$/$344\times3$& 1342259618/19 & 2013-01-14/15\\
W1422+29 & 14 22 30.34 & +29 52 24.2   &0.1133 & \OI/\CII &$384\times10\times2$/$344\times3$& 1342262004/05 & 2013-01-24\\
W1519+52 & 15 19 07.33 & +52 06 06.1  &0.1378 & \OI/\CII &$384\times10\times1$/$344\times5$& 1342262538/39 & 2013-01-29\\
W1634+20 & 16 34 59.83 & +20 49 36.1  &0.1285 & \OI/\CII &$384\times10\times2$/$344\times4$& 1342263467/66 & 2013-02-10\\
W1659+18{\tablenotemark{c}} & 16 59 39.77 & +18 34 36.8  &0.1709 & \OI/\CII &$384\times10\times2$/$344\times2$& 1342263468/38910 & 2013/12-02-10\\
\enddata
\tablecomments{Column (1): source; Columns (2) and (3): right ascension and declination; Column (4): optical redshift from SDSS DR 10; Column (5): targeted line for the corresponding  observation ID; Column (6): exposure time; Column (7): observation ID; Column (8): observation date.}
\tablenotetext{a}{At the same time, a different spectrum window was observed for the continuum, as shown in Fig. 1.}
\tablenotetext{b}{${\rm Time\ per\ repetition} \times {\rm number\ of\ repetitions\ per\ nod\ cycle}\,(\times {\rm number\ of\ nod\ cycles)}$. Only the science time is shown, and the on-source exposure time is a half of the listed number.}
\tablenotetext{c}{Observations for the \CII\ was performed by the program OT1\_dweedman\_1 (PI: D. Weedman).}
\end{deluxetable*}

The PACS spectroscopic data for all but one source (W1659+18; see Table \ref{table:obslog}), whose \CII\ observation was performed by the program OT1\_dweedman\_1 (PI: D. Weedman), were collected as the OT2 program (ID: OT2\_lyan\_5, PI: L. Yan) awarded with about 24\,hours of observing time. The observations were carried out between 2012 September and 2013 February (see Table \ref{table:obslog}) using the Integral Field Spectrometer (IFS). The IFS on PACS can perform simultaneous spectroscopic observations in the 51\,-\,73 or 71\,-\,105\,\mum\ and the 103\,-\,220\,\mum\ ranges. The integral field unit (IFU) is composed of a 5$\times$5 array of individual detectors (spaxels), which has a projected field of view (FoV) of 47$^{\prime\prime}\times$47$^{\prime\prime}$ on the sky. The point-spread function (PSF) of full width at half-maximum (FWHM) for the PACS IFS is $\sim$9$^{\prime\prime}.5$ between 60\,\mum\ and $\sim$\,110\,\mum, and increases to $\sim$14$^{\prime\prime}$ by 200\,\mum. The spectral resolution is around 55\,\kms\ (the third grating order; $\sim$160\,\kms\ for the second grating order) and 210\,\kms\ for \OIFIR\ and \CII\ respectively, allowing us to resolve the \OI\ and most \CII\ lines in velocity (see below).

At $z$$\sim$0.1\,-\,0.2, an individual PACS spaxel corresponds to $\sim$19\,-\,32\,kpc, and thus the single pointed observation mode was used for all sources in our sample, with targets placed in the central spaxel. The \OIFIR\ and \CII158\,\mum\ lines (Table \ref{table:obslog}) were observed separately using the line spectroscopy mode. For each line observation, an additional spectral window (red/blue for \OI/\CII) was observed simultaneously for the continuum measurement. We used the standard chopping-nodding observing mode, in which the source is observed by alternating between the on-source position and a clean off-source position, and adopted the chopper throw of ``Large" ($6^\prime$; except for the \CII\ observation of W1659+18, in which a medium throw of $3^\prime$ was used) in order to avoid possible contamination of the background from nearby sources. To eliminate the telescope background emission, two nod positions were used.

The PACS data reduction was performed in the Herschel Interactive Processing Environment (HIPE) version 12.1 (Ott 2010) using a customized chop/nod background normalization pipeline script, which is particularly aimed at fainter sources with long duration observations. The final spectra were extracted from the central spaxel and corrected for aperture loses using a point source PSF. All of our sources are unresolved at 22\,\mum\ in the \wise\ images. Since 22\,\mum\ \wise\ images have an angular resolution of $\sim$12$^{\prime\prime}$, similar to that of PACS at 170\,\mum, and dust emissions at 20\,-\,160\,\mum\ generally have similar concentrations (e.g. Mu\~{n}oz-Mateos et al. 2009), we do not expect significant extended emission outside the adopted aperture. Indeed, we found that the summed spectra from the central $3\times3$ spaxels are generally consistent with those extracted from the central spaxel (both after aperture correction), but much noisier due to the different continuum slopes. The PACS spectra for our 9 targets are shown in Figure \ref{Fig1}.

{\renewcommand{\arraystretch}{1.3}
\begin{deluxetable*}{llcccccccc}[pthb]
\tablecaption{Measured Line Fluxes\label{tableline}}
\tablewidth{0.75\textwidth}
\tabletypesize{\scriptsize}
\tablehead{
\colhead{\multirow{2}{*}{Galaxy}}&\multicolumn{3}{c}{\OI\ 63 \mum}&\multicolumn{3}{c}{\CII\ 158 \mum}&\multicolumn{3}{c}{\OI\ 6300 \AA}\\
\cline{2-4}
\cline{5-7}
\cline{8-10}
&\colhead{flux}&\colhead{FWHM}&\colhead{redshift}&\colhead{flux}&\colhead{FWHM}&\colhead{readshift}&\colhead{flux}&\colhead{FWHM}&\colhead{readshift}
}
\startdata
W0752+19  & 4.37$\pm$0.34 & 232 &0.1172& 3.14$\pm$0.27   &  198&0.1171&624.0$\pm$6.2&372&0.1171\\
W0752+19 &  11.0$\pm$1.4 & 724 & 0.1169& $<$1.10\tablenotemark{c} &\nodata  &\nodata&1380$\pm$3.1&1335&0.1167\\
W0946+13 & $<$0.46&\nodata &\nodata&$0.88\pm0.18$&311&0.1331&235.2$\pm$5.3&319&0.1330\\
W0946+13 &\nodata&\nodata &\nodata&\nodata&\nodata&\nodata&825.3$\pm$2.6&1823&0.1336\\
W1005+43 & $1.41\pm0.13$ & 318 &0.1793 &$0.93\pm0.30$ & 187 &0.1794&$<$26.9&\nodata&\nodata \\
W1117+44 &$<$0.67 &\nodata &\nodata & $<$0.68 &\nodata &\nodata&$<90.2$&\nodata&\nodata\\
W1355+20 &$1.48\pm0.30$ &664 &0.1976&$1.88\pm0.41$ &262 &0.1975&$<$35.0&\nodata&\nodata \\
W1355+20 &\nodata &\nodata &\nodata &$3.95\pm1.18$ &676 &0.1945&\nodata&\nodata&\nodata\\
W1422+29 &$2.97\pm0.33$ &276&0.1141&$2.22\pm0.21$&264&0.1141&37.0$\pm$4.4&147&0.1143\\
W1422+29 &\nodata&\nodata&\nodata&\nodata&\nodata&\nodata&313.3$\pm$2.4&924&0.1139\\
W1519+52{\tablenotemark{a}} &\nodata &\nodata&\nodata&$7.68\pm0.51$&474&0.1386&20.1$\pm$4.6&143&0.1382\\
W1519+52 &\nodata &\nodata&\nodata&\nodata&\nodata&\nodata&94.2$\pm$3.3&692&0.1385\\
W1634+20 &$1.81\pm0.30$ &230&0.1293 &$2.80\pm0.27$ &295&0.1293&32.1$\pm$2.4&362&0.1291\\
W1634+20{\tablenotemark{b}} &\nodata&\nodata&\nodata&\nodata&\nodata&\nodata&151.9$\pm$5.3&3022&0.1304\\
W1659+18 &$2.79\pm0.34$ & 487&0.1711 &$2.18\pm0.22$ &441&0.1709&157.9$\pm$1.5&608&0.1711\\
\enddata
\tablecomments{Fluxes are in units of 10$^{-17}$ W m$^{-2}$, and FWHM in units of km s$^{-1}$. The second line of each source, if exists, gives the information of the second component of the emission.}
\tablenotetext{a}{The observed spectra do not cover the expected wavelength of the \OI\ line for this source.}
\tablenotetext{b}{This component suffers a large uncertainty due to the worse quality of the spectra within this wavelength range, and thus is not taken into account in our analysis.}
\tablenotetext{c}{This flux limit was calculated by assuming that the \CII\ has the same line width as the \OI\ broad component.}
\end{deluxetable*}   

\setcounter{figure}{1}
\begin{subfigures}
\begin{figure*}[pthb]
\centering
\includegraphics[width=0.99\textwidth,bb=138 276 490 512]{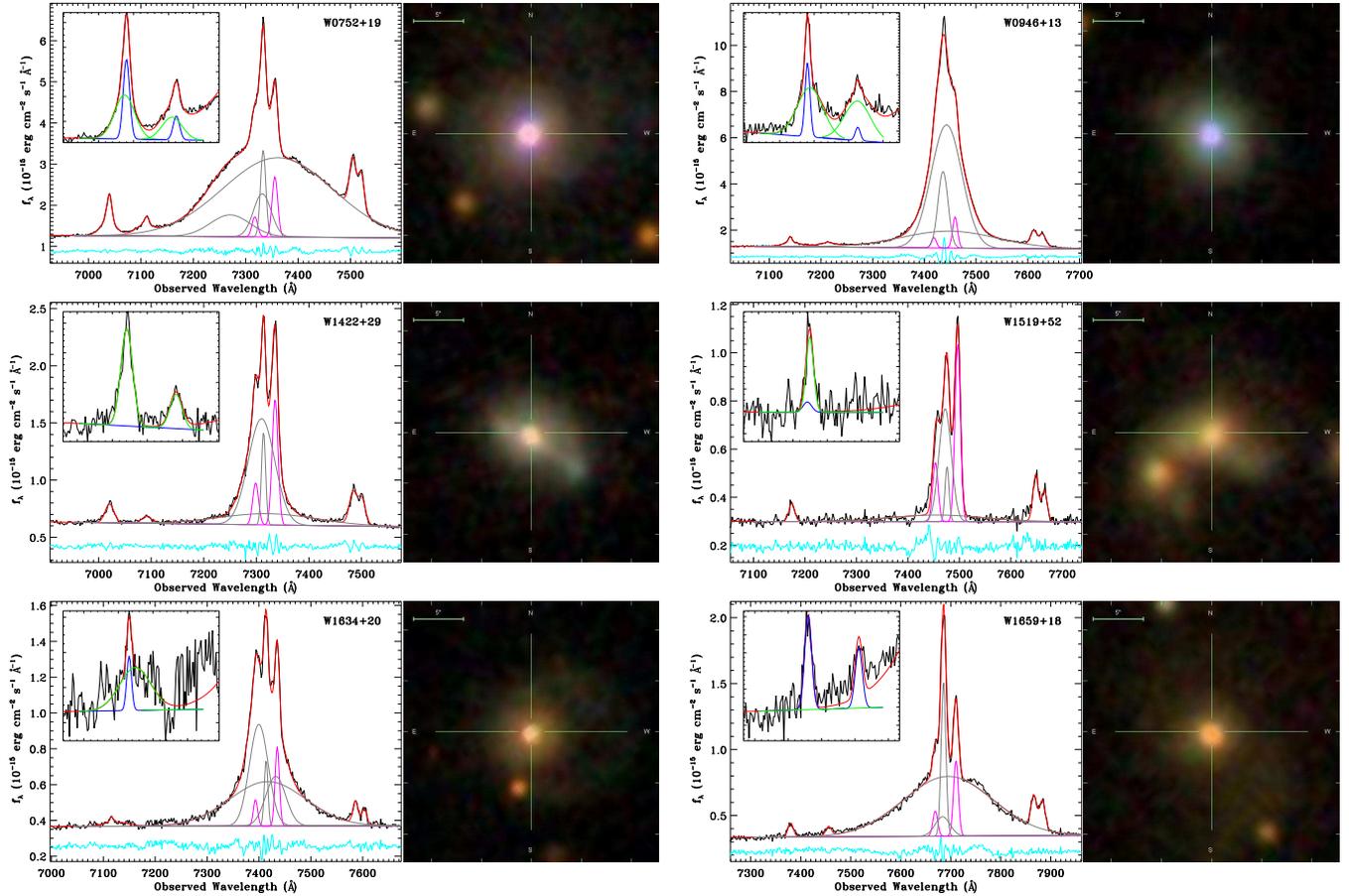}
\caption{SDSS optical spectra around the \Ha\ line, and false color images (R: $i$-band, G: $r$-band, B: $g$-band) of the QSOs in our sample with at least one of the \OIDOUBLE\ doublet detected. For each source, the observed and best fitted spectra are shown by the black and red lines respectively, with the cyan line (arbitrarily shifted vertically) giving the residuals. Individual components of the H$\alpha$ line and the \NIIDOUBLE\ doublet (plus the pseudo-continuum) are shown by the gray and magnetar lines respectively.  The inset in each spectrum panel presents a enlarged view of the \OIDOUBLE\ emissions, with the narrow and broad components (plus the pseudo-continuum) shown by the blue and green lines, respectively.}
\label{figsdssa}
\end{figure*}

\setcounter{figure}{1}
\begin{figure}[ptbh]
\centering
\includegraphics[width=0.47\textwidth,bb=77 270 312 592]{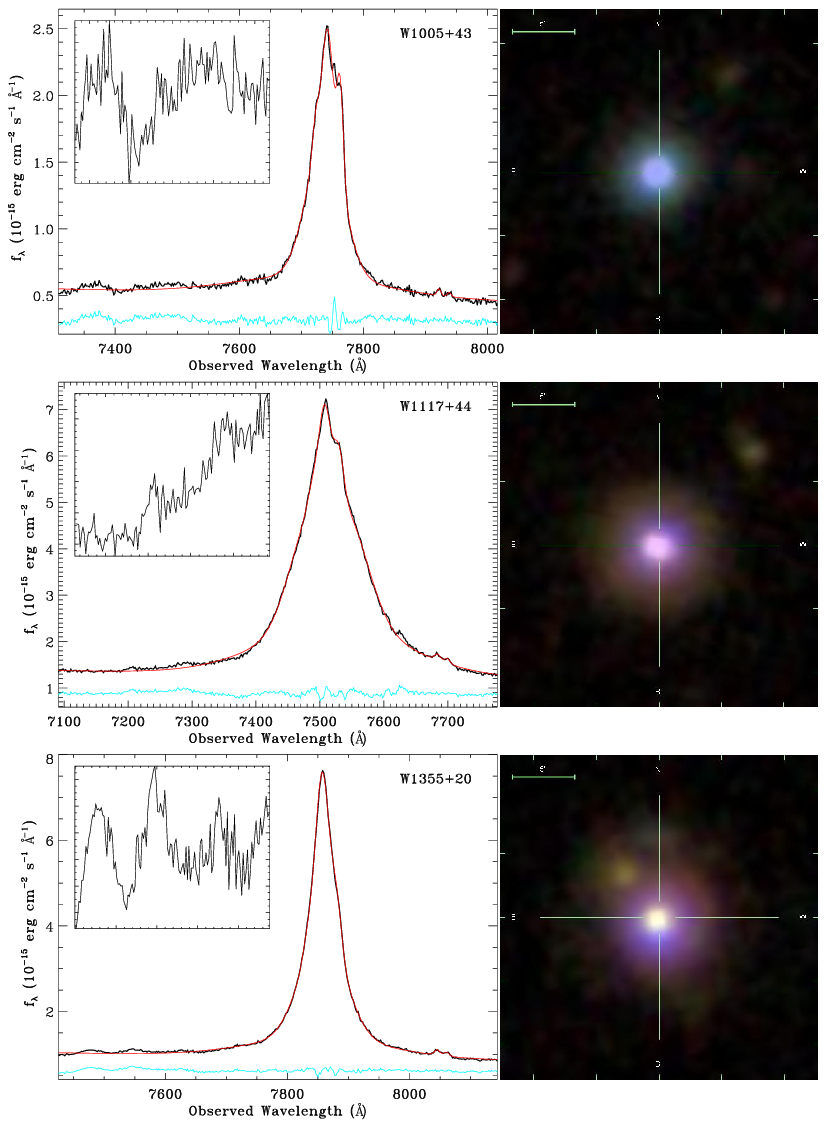}
\caption{Same as Fig. 2A, except for that the \OIDOUBLE\ emissions only show very weak features or non-detections.}
\label{figsdssb}
\end{figure}
\end{subfigures}

\begin{deluxetable}{lcccc}[pth]
\tablecaption{PACS Spectrophotometric Results\label{tablephotpacs}}
\tablewidth{0.47\textwidth}
\tabletypesize{\scriptsize}
\tablehead{
\colhead{\multirow{2}{*}{Galaxy}}&\colhead{PACS1}&\colhead{PACS2}&\colhead{PACS3}&\colhead{PACS4}\\
&\colhead{$\lambda_1$}&\colhead{$\lambda_2$}&\colhead{$\lambda_3$}&\colhead{$\lambda_4$}
}
\startdata
W0752+19&$<$318&$<$744&$<$252&$<$201\\
&70.60&88.15&176.27&211.76\\
W0946+13&$<$474&$<$552&$<$282&$<$156\\
&71.61&89.42&178.83&214.81\\
W1005+43&$  738\pm   43$&$  983\pm  198$&$  461\pm   45$&$<$702\\
&74.50&92.99&148.98&185.95\\
W1117+44&$<$624&$<$735&$<$582&$<$231\\
&72.27&90.25&180.48&216.78\\
W1355+20&$  252\pm   84$&$<$879&$  138\pm   35$&$<$660\\
&75.62&94.38&151.21&188.71\\
W1422+29&$ 1182\pm  153$&$ 1105\pm  213$&$  468\pm  148$&$  912\pm  105${\tablenotemark{a}}\\
&70.36&87.85&175.68&211.04\\
W1519+52&$  852\pm  240$&$ 1048\pm  239$&$  820\pm  167$&$ 1103\pm  164${\tablenotemark{a}}\\
&71.61&89.78&179.54&214.83\\
W1634+20&$  642\pm  121$&$  840\pm  217$&$  397\pm  122$&$  728\pm   61${\tablenotemark{a}}\\
&71.34&89.10&178.24&214.03\\
W1659+18&$317\pm71$&$<$891&$  268\pm   26$&$<$864\\
&74.02&92.38&148.01&184.74\\
\enddata
\tablecomments{For each source, the first line lists the fluxes, and the second line gives the corresponding wavelengths (observed frame). Fluxes are in units of mJy, and wavelength in units of \mum.} 
\tablenotetext{a}{The observed spectra with $\lambda>190\,\mu{\rm m}$ are in the leakage region and contaminated by spectral features from $95-110$ \mum. Hence the measured flux should be overestimated.}
\end{deluxetable} 

We fitted the observed profile and continuum simultaneously using a single- or two-component Gaussian plus a 1st order polynomial. The fitting results are illustrated in Figure \ref{Fig1} using (green) solid lines.  Table~\ref{tableline} lists the measured line fluxes as well as the intrinsic line widths, which have been corrected for the instrumental spectral resolution. We used $\sigma_{\rm true} = \sqrt{\sigma_{\rm obs}^2 - \sigma_{\rm inst}^2}$ to make this correction. If a line is not detected, we list its upper limit, which was calculated as $3\sigma$ times the spectral resolution (FWHM of the instrumental profile) at the expected wavelength of the line.

The continuum flux was estimated to be the median value of the spectrum (after removing the fitted line profile for the line-emission window), and the error ($\sigma$) was calculated using 1.48$\times$${\rm MAD}$ assuming normally distributed noise, where MAD is defined as the median value of the absolute deviations from the median data. Continuum fluxes are listed in Table \ref{tablephotpacs}. For the continuum flux below our sensitivity, we give the $3\sigma$ flux as upper limit. We note that the observed spectra with $\lambda$\,$=$\,190\,-\,220\,$\mu{\rm m}$ are in the leakage region and contaminated by spectral features from 95\,-\,110\,\mum. Hence the measured continuum flux within this wavelength range could be over-estimated by a factor of three{\footnote{See \url{https://nhscsci.ipac.caltech.edu/workshop/DP\_Workshop
\_Aug\_2013/PACS/presentations/pipeline.pdf}}}.

\subsection{Optical Spectroscopy}
The analyses in this paper use the QSO optical spectra taken by {\it SDSS}. These spectra provide information on \OIDOUBLE\ emission lines, which are in complementary to the FIR \OIFIR\ lines observed by {\it Herschel}/PACS.  The reduced, calibrated one-dimensional {\it SDSS} spectra, as well as the three-color RGB images (based on the $g$, $r$ and $i$ images; Lupton et al. 2004), are obtained from the {\it SDSS} Date Release 10 (DR10; Ahn et al. 2014) via the Science Archive Server{\footnote{\url{http://dr10.sdss3.org}}}. In Figures \ref{figsdssa} and \ref{figsdssb}, we show the {\it SDSS} spectra and RGB images for our sample objects with detections and non-detections of the \OIDOUBLE\ emissions, respectively.

To obtain the fluxes of the \OIDOUBLE\ and \Ha\ lines, we fitted the continuum, \OIDOUBLE, \Ha\, \NIIDOUBLE\ and \SIIDOUBLE\ simultaneously. During the fitting process, we adopted Gaussian profiles for line emissions, and up to 2 components (narrow and broad) for the \OIDOUBLE\ lines, and up to 4 components (1 narrow and 3 broad) for the \Ha\ line. As shown in Shen et al. (2011), a multi-component fitting to the H$\alpha$ line generally gives a better fitted result. The ratio of the \NIIDOUBLE\ doublet is fixed as 1/3. We also assumed that each doublet have the same line width. The best fitted results (red lines) and residuals (arbitrarily shifted vertically; cyan lines) are also superposed in Figures \ref{figsdssa} and \ref{figsdssb}. For a clearer visualization of the \OIDOUBLE\ doublet, the zoomed spectrum of 6220\,-\,6420\AA\ (rest frame) are plotted in the inset of each panel, with the fitted (if have) individual component(s) overlaid. We also used the similar method to fit the H$\beta$ and \OIIIDOUBLE\ lines.

\section{Results and Discussion}
\subsection{IR Spectral Energy Distribution}
To estimate the IR/FIR properties of our sample QSOs, such as the total infrared luminosity (\LIR; 8\,-\,1000 \mum), FIR luminosity (\LFIR; 42.5\,-\,122.5\,\mum) and the fractional contributions to the total \LFIR\ from starburst activities ($f_{\rm SB}$), we also compiled near-IR to FIR photometries from 2MASS, \wise\ and \iras\ catalogs, and list the results in Table \ref{tablephotlit}. Combined with our own PACS spectrophotometric results, these data allow us to obtain the aforementioned parameters (\LFIR, \LIR) in a uniform manner by fitting the observed IR SEDs.

To fit the observed IR SEDs, we adopted a simple two-component method, i.e. by combining a starburst (SB) component and a QSO component. For each component, we allow its fractional contribution to the total SED to vary from 0 (thus only one component) to 1. The SED templates were adopted from the SWIRE library (Polletta et al. 2007), which include 6 SB (NGC\,6090, NGC\,6240, M\,82, Arp\,220, IRAS\,22491-1808, and IRAS\,20551-4250) and 5 QSO. Amongst these QSO templates, three of them (QSO1, TQSO1, and BQSO1) represent optically-selected QSOs with different values of infrared/optical flux ratios, and two are type 2 QSOs (QSO2 and Torus).

The final algorithm we used to fit the SEDs is similar to that proposed in Sawicki (2012), which properly treat non-detections. The equation A10 in Sawicki (2012) is used to calculate the $\chi^2$, namely,
\begin{equation}
\begin{split}
\chi^2  = & \sum \limits_{i} \left(\frac{f_{{\rm o},i}-f_s f_{{\rm m},i}}{\sigma_i}\right)^2 \\ & - 2\sum \limits_{j} \ln \left\{ \sqrt{\frac{\pi}{2}}\sigma_j \left[ 1+ {\rm erf} \left(\frac{f_{{\rm lim},j}-f_s f_{{\rm m},j}}{\sqrt{2}\sigma_j}\right)\right]\right\}.
\end{split}
\end{equation}
where $f_{{\rm o},i}$ refers to the observed flux density of the detected band $i$, $f_{{\rm lim},j}$ the 1-$\sigma$ flux detection threshold in the $j$th band, $f_{\rm m}$ the model flux, $\sigma$ the uncertainty, $f_s$ the scaling factor between the observed data and the model, and ${\rm erf}(x)$ the error function.

Figure \ref{Figseds} shows the observed data points and best-fit SEDs for our sample QSOs. In each panel, we plotted SB (dashed line) and QSO (dotted line) components, and labeled the name(s) of the best fitted template(s). The gray lines in each panel give all model results at the 95\% confidence level. The combination of the two types of templates fits our observed SEDs generally well. The fitted parameter space is better constrained by having detections at the FIR bands. For three objects (W1422+29, W1519+52 and W1634+20), the data points at the longest wavelength fall in the leakage region and can be over-estimated. Thus we discarded these points during the fitting process. Table \ref{tableratio} lists the derived parameters, such as \LIR, \LFIR, $f_{\rm SB}$ and FIR color (\fircolor) from the best-fit SEDs.

\begin{turnpage}
\begin{deluxetable*}{lccccccccccc}[ptbh]
\tablecaption{Photometry\label{tablephotlit}}
\tablewidth{0pt}
\tabletypesize{\scriptsize}
\tablehead{
\colhead{Galaxy}&\colhead{$f_{1.2}$}&\colhead{$f_{1.6}$}&\colhead{$f_{2.2}$}&\colhead{$f_{W1}$}&\colhead{$f_{W2}$}&\colhead{$f_{W3}$}&\colhead{$f_{W4}$}&\colhead{$f_{12}$}&\colhead{$f_{25}$}&\colhead{$f_{60}$}&\colhead{$f_{100}$}\\
\colhead{(1)}&\colhead{(2)}&\colhead{(3)}&\colhead{(4)}&\colhead{(5)}&\colhead{(6)}&\colhead{(7)}&\colhead{(8)}&\colhead{(9)}&\colhead{(10)}&\colhead{(11)}&\colhead{(12)}
}
\startdata
W0752+19&$    3.66\pm    0.27$&$    3.52\pm    0.41$&$    6.05\pm    0.48$&$   10.76\pm    0.39$&$   17.37\pm    0.58$&$   48.6\pm    5.7$&$  160\pm   19$&\nodata&\nodata&\nodata&\nodata\\
W0946+13&$    3.09\pm    0.16$&$    3.87\pm    0.22$&$    6.26\pm    0.29$&$    9.95\pm    0.36$&$   14.91\pm    0.49$&$   45.9\pm    5.3$&$  140\pm   18$&\nodata&\nodata&\nodata&\nodata\\
W1005+43&$    1.84\pm    0.11$&$    2.59\pm    0.16$&$    5.56\pm    0.26$&$    7.24\pm    0.27$&$   11.77\pm    0.41$&$   49.4\pm    5.7$&$  131\pm   16$&$<$233.6&$  185\pm   19$&$  558\pm   45$&$  874\pm  131$\\
W1117+44&$    3.38\pm    0.16$&$    3.97\pm    0.19$&$    7.76\pm    0.31$&$   13.91\pm    0.50$&$   22.90\pm    0.76$&$   58.8\pm    6.7$&$  148\pm   18$&$  110\pm   37$&$  149\pm   33$&$  191\pm   47$&$  200\pm   60$\\
W1355+20&$    2.73\pm    0.17$&$    4.32\pm    0.28$&$    8.13\pm    0.39$&$   16.62\pm    0.60$&$   24.99\pm    0.85$&$   73.6\pm    8.5$&$  171\pm   21$&$<$341&$  210\pm   30$&$  264\pm   43$&$<$985\\
W1422+29&$    3.30\pm    0.28$&$    3.56\pm    0.38$&$    3.84\pm    0.40$&$    2.35\pm    0.09$&$    2.69\pm    0.09$&$   28.3\pm    3.3$&$  132\pm   16$&$<$203&$  165\pm   18$&$  960\pm   77$&$ 1390\pm  194$\\
W1519+52&$    1.99\pm    0.13$&$    3.61\pm    0.24$&$    6.91\pm    0.34$&$   11.84\pm    0.42$&$   19.74\pm    0.66$&$   75.9\pm    8.7$&$  220\pm   26$&$   79\pm   18$&$  279\pm   22$&$  780\pm   47$&$ 1341\pm  134$\\
W1634+20&$    3.18\pm    0.35$&$    5.20\pm    0.55$&$   10.20\pm    0.74$&$   15.54\pm    0.56$&$   22.09\pm    0.76$&$   46.0\pm    5.3$&$  119\pm   14$&$<$155&$  141 \pm   11$&$  559\pm   45$&$ 1170\pm  199$\\
W1659+18&$    1.49\pm    0.09$&$    2.31\pm    0.13$&$    4.48\pm    0.22$&$   12.52\pm    0.46$&$   18.82\pm    0.63$&$   56.2\pm6.4$&$  149\pm   18$&$<$218&$163\pm   18$&$277\pm42$&$<$1560\\
\enddata
\tablecomments{Column (1): source; Columns $(2)-(4)$: fluxes of the 2MASS $J$-, $H$- and $K_s$-band, respectively; Columns $(5)-(8)$: fluxes of the four \wise\ bands; Columns $(9)-(12)$: fluxes of the four \iras\ bands. Fluxes are in units of mJy.}
\end{deluxetable*} 
\end{turnpage}

\begin{figure*}[ptbh]
\centering
\includegraphics[width=0.95\textwidth,bb=81 397 732 1036]{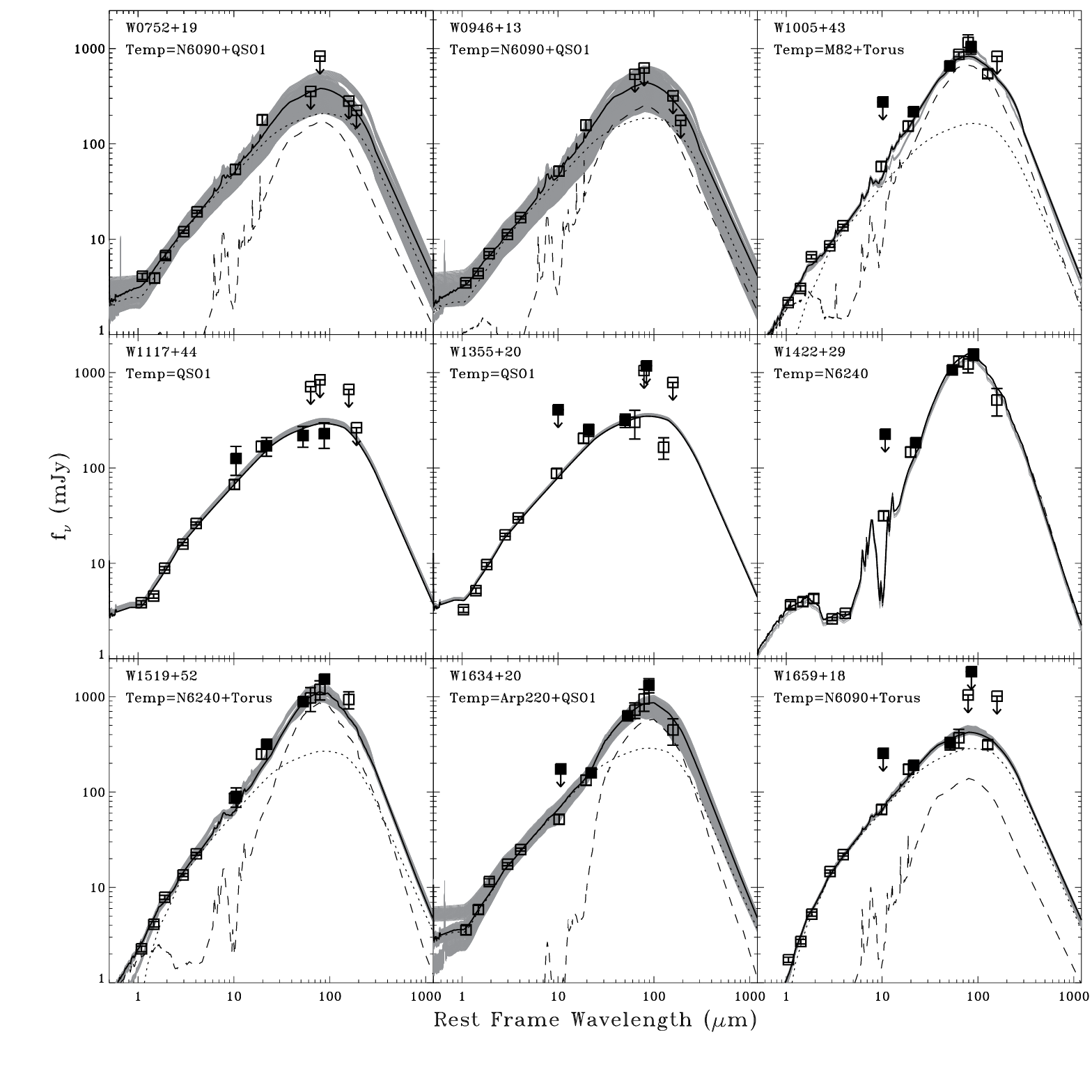}
\caption{The NIR to FIR SEDs (as listed in Tables \ref{tablephotpacs} and \ref{tablephotlit}) for our sample QSOs. The \iras\ photometry is plotted as solid squares, and arrows indicate $3\sigma$ upper limits. The best fitted result is shown by the thick solid line, with individual components from starburst and AGN templates (as labelled in each panel) of the SWIRE library (Polletta et al. 2007) plotted as dashed and dotted lines respectively. The gray lines give all model results at the 95\% confidence level.}
\label{Figseds}
\end{figure*}

\subsection{Our sample of IR Bright QSOs are mostly composite systems}
The $L_{\rm IR}$ values derived from the SED fitting are in the range of 4.6--18.6$\times 10^{11} L_{\sun}$ for our sample (Table \ref{tableratio}), similar to (U)LIRGs. The difference from local ULIRG galaxies is that these IR bright QSOs have very warm FIR colors with $f_{25}/f_{60}$ ratios of $0.14-0.63$ (median value of 0.41), typical values seen amongst {\it IRAS} discovered IR bright QSO/AGNs.  The $f_{60}/f_{100}$ FIR colors are very similar, $\sim$0.93, as shown in Table \ref{tableratio}. We evaluate our full sample with regard to the starburst component contribution fraction $f_{\rm SB}$, and find that it varies from 0 (i.e. complete QSO SED) to 1 (complete SB SED) in the sample. We found 4 systems are AGN dominated with $f_{\rm SB} <0.5$, and the remaining 5 QSOs are starburst dominated with $f_{\rm SB}>0.5$. The median $f_{\rm SB} = 0.55$ indicates both starburst and QSO contributions to their IR luminosities. We note that our QSOs have a factor of (5-10) less far-IR emission (\LFIR\ $\sim$ a few times of $10^{11}L_\odot$) than that of local ULIRGs. This is expected since these sources are not very bright in {\it IRAS} catalogs. 

The nature of SB-QSO composite SEDs in our sample is consistent with the optical morphologies, as shown in Figures \ref{figsdssa} and \ref{figsdssb}, where {\it SDSS} color images clearly reveal the extended emissions from host galaxies. Bright IR emissions in excess of a pure QSO template in these sources could be due to dust obscured star formation in their host galaxies. This is consistent with the fact that the QSO bolometric luminosities, derived from 5100\,\AA\ luminosities (Shen et al. 2011), are almost all less than our SED derived total IR luminosities, $L^{\rm QSO}_{\rm bol} < L_{\rm IR}$.  Monochromatic luminosity at 5100\,\AA\ is   usually measured from an optical spectrum, which is dominated by the light from the central accreting black hole (QSO), therefore, the derived $L^{\rm QSO}_{\rm bol}$  represents mostly the QSO, but not its host galaxy.  

\renewcommand{\arraystretch}{1.0}
\begin{center}
\begin{deluxetable*}{lccccccccccc}[pthb]
\tablecaption{Derived Properties\label{tableratio}}
\tablewidth{0pt}
\tabletypesize{\scriptsize}
\tablehead{
\colhead{\multirow{2}{*}{Source}}&\colhead{$L_{\rm bol}^{\rm QSO}$}&\colhead{$M_{\rm BH}$}&\colhead{$L_{\rm IR}$}& $L_{\rm FIR}$ &\multirow{2}{*}{$f_{\rm SB}$}&\multirow{2}{*}{$\frac{f_{60}}{f_{100}}$}&\multirow{2}{*}{$\frac{{[{\rm O}\,{\tiny {\textsc{i}}}]}}{{\rm FIR}}$}&\multirow{2}{*}{$\frac{{[{\rm C}\,{\tiny {\textsc{ii}}}]}}{{\rm FIR}}$}&\multirow{2}{*}{$\frac{{[{\rm O}\,{\tiny {\textsc{i}}}]}}{{[{\rm C}\,{\tiny {\textsc{ii}}}]}}$}&\colhead{$R_{\rm ENLR}$}&\colhead{$M_{\rm dyn} \sin^2i$}\\
&\colhead{(\lsolar)}&\colhead{($M_\odot$)}&\colhead{(\lsolar)}&\colhead{(\lsolar)}&&&&&&\colhead{(kpc)}&\colhead{($M_\odot$)}\\
\colhead{(1)}&\colhead{(2)}&\colhead{(3)}&\colhead{(4)}&\colhead{(5)}&\colhead{(6)}&\colhead{(7)}&\colhead{(8)}&\colhead{(9)}&\colhead{(10)}&\colhead{(11)}&\colhead{(12)}
}
\startdata
W0752\tablenotemark{a} &11.87&9.43&11.67& 11.16 &0.43& 0.94&$-2.01$&$-2.70$&4.89&3.2&11.6\\
W0752\tablenotemark{a} &---&---&---& --- &---& ---&$-2.56$&$-2.70$&1.39&---&---\\
W0752\tablenotemark{a} &---&---&---& ---&---& ---&$-2.15$&$<$$-3.15$&$>$10.0&---&---\\
W0946 &12.08&8.39&11.80&11.33&0.55&0.95&$<$$-3.59$&$-3.31$&$<$$0.52$&4.0&11.0\\
W1005 &11.92&7.94&12.27& 11.88 &0.79&0.97 &$-3.37$&$-3.55$ &1.51 &3.4&10.9\\
W1117 &12.13&8.77&11.85&11.26 &0.0 &0.93&$<$$-3.27$&$<$$-3.26$&---&4.3&---\\
W1355 &12.30&8.21&12.23&11.63&0.0&0.94&$-3.00$&$-2.90${\tablenotemark{b}} &0.79{\tablenotemark{b}}&5.1&11.7\\
W1422 &11.35&7.52&11.94&11.71&1.0&0.96&$-3.29$&$-3.42$&1.34&1.9&10.5\\
W1519 &11.12&8.07&12.12&11.76&0.74&0.95&\nodata&$-2.75$&\nodata&1.5&10.4\\
W1634 &11.07&8.54&11.96&11.56&0.63&0.86 &$-3.26$&$-3.07$&0.65&1.4&10.9\\
W1659 &11.46&9.16&12.10&11.57&0.31&0.95 &$-2.80$&$-2.91$&1.28&2.1&11.1\\
\enddata
\tablecomments{Column 1: abbreviated source name; Column 2: bolometric luminosity of the QSO adopted from Shen et al. (2011), and have been corrected for the host contamination using their equation (1); Column 3: black hole mass adopted from Shen et al. (2011); Columns 4 and 5: total IR and FIR luminosity obtained from the SED fitting, respectively; Column 6: fractional contribution to $L_{\rm FIR}$ from the starburst component based on the SED-fitting results; Column 7: 60-to-100 \mum\ flux ratio obtained through SED fit; Columns 8 and 9: logarithm of the \OI- and \CII-to-FIR ratios, respectively; Column 10: \OI-to-\CII\ ratio; Column 11: size of the extended narrow-line region estimated using the continuum luminosity at 5100 \AA\ ($L_{5100\mathring{\mathrm{A}}}$) and equation (3) in Husemann et al. (2014); Column 12: dynamical mass calculated using $R_{\rm ENLR}$; Column 13: size of the broad line region estimated using $L_{5100}$ and the best fitted results in Bentz et al. (2009); Column 14: dynamical mass calculated using $R_{\rm BLR}$. {\bf All luminosities, masses, and line-to-FIR ratios are in logarithmic scale.}} 
\tablenotetext{a}{The ratios listed in the first, second and third line were calculated with line fluxes from the total, narrow, and broad component, respectively.}
\tablenotetext{b}{For the \CII\ emission, only the component with similar velocity to \OI\ line was used.}
\end{deluxetable*}              
\end{center}

\begin{figure*}[pthb]
\centering
\includegraphics[width=0.99\textwidth,bb=1 40 500 290]{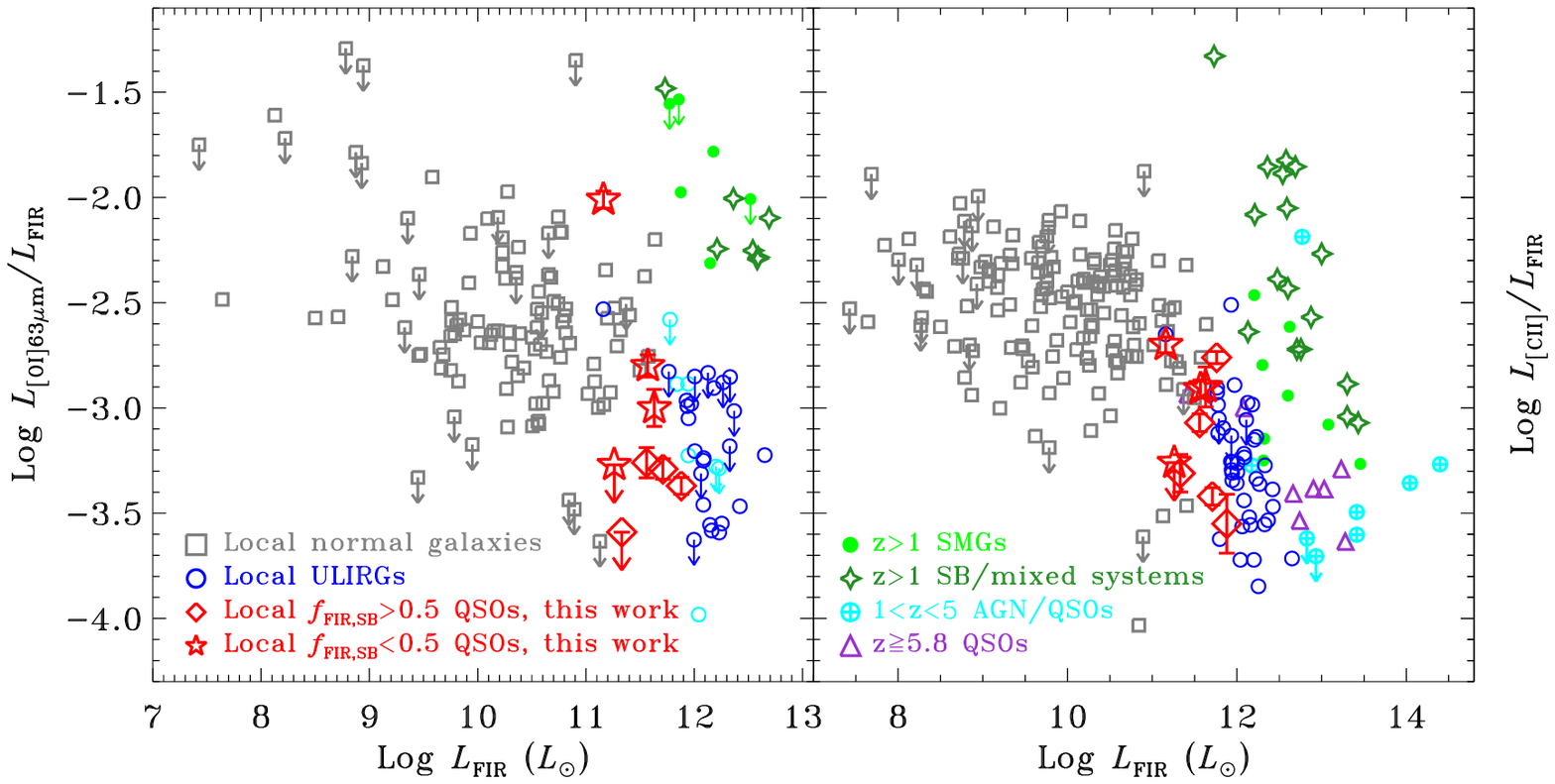}
\caption{Left: \LOI/\LFIR\ vs. \LFIR. Right: \LCII/\LFIR\ vs. \LFIR. Red symbols show QSOs in current work, with open diamonds and five-angle stars representing $f_{\rm SB} > 0.5$ and $f_{\rm SB} < 0.5$ respectively. To compare our data with other samples, we also plot local normal galaxies (including LIRGs; Brauher et al. 2008), local ULIRGs (Brauher et al. 2008; Farrah et al. 2013), $z>1$ star-forming (SB)/mixed systems (Stacey et al. 2010; Brisbin et al. 2015), $z>2$ submillimeter bright galaxies (SMGs; Maiolino et al. 2009; Ivison et al. 2010; De Breuck et al. 2011; Valtchanov et al. 2011; Coppin et al. 2012; Swinbank et al. 2012; Wagg et al. 2012; Riechers et al. 2013), $1 < z <5$ \CII\ detected AGN/QSOs (Pety et al. 2004; Stacey et al. 2010; Gallerani et al. 2012; Wagg et al. 2012), and $z>5.8$ QSOs (Maiolino et al. 2005; Venemans et al. 2012; Leipski et al. 2013; Wang et al. 2013; Willott et al. 2013).}
\label{Figlineratio}
\end{figure*}

\subsection{Physical Characteristics of the ISM in Our IR Bright QSOs}

The \OI\,63$\mu$m emission arises exclusively from the neutral phase of gas clouds because atomic oxygen has an ionization potential just above 13.6 eV.  Although \CII\ line is considered a primary tracer of photo-dominated regions (PDRs; Tielens \& Hollenbach 1985), it can also come from diffused, ionized gas as it only takes $\sim$11.3\,eV to form C$^+$. Neutral and ionized gas can be heated by photoelectric process (PE; Tielens \& Hollenbach 1985), then cooled via collisional excitation of C$^+$, O and other elements. The \OI\,63$\mu$m emission has a higher excitation temperature ($\sim$228 K) and considerably higher critical density (in this paper, all densities are for hydrogen nuclei, not electrons, unless otherwise stated) ($n_{\rm cr}$$\sim$5$\times10^5$\,cm$^{-3}$) than that of  \CII\ emission (92\,K and $\sim$3$\times10^3$\,cm$^{-3}$ respectively).  The cooling rate per volume is $\propto n_{\rm gas}^2$ when $n_{\rm gas}<n_{\rm cr}$, and $\propto n_{\rm gas}$ when $n_{\rm gas}>n_{\rm cr}$. On the other hand, the heating rate per volume always increases faster than $n_{\rm gas}$ so that temperature is higher at higher densities. Therefore, the combination of the \CII\,158\,\mum\ and \OIFIR\ lines allows unique determination of the physical properties (e.g. temperature and density) of PDR gas (Kaufman et al. 1999). Furthermore, the ratio of the combined luminosity of these two lines to the IR continuum can be used as a proxy for the PE heating efficiency (e.g. Tielens \& Hollenbach 1985), which represents the FUV energy that goes into to the gas, divided by the far-ultraviolet (FUV) energy deposited in dust grains.

\subsubsection{\LOI-to-\LFIR\ and \LCII-to-\LFIR\ Ratios}

Table \ref{tableratio} lists the derived line luminosities (\LOI\ and \LCII, respectively) and line-to-\LFIR\ ratios for our sample. The \LOI/\LFIR\ values are in the range of 2.6$\times$$10^{-4}$ and $10^{-2}$ with a median value of 5.2$\times10^{-4}$, including the two upper limits. The \LCII/\LFIR\ ratios span between 2.8$\times$$10^{-4}$ and 2.0$\times$$10^{-3}$ with a median of 7.6$\times$$10^{-4}$.  Figure~\ref{Figlineratio} plots the \OI\ (left panel) and \CII\ (right panel) to FIR luminosity ratios as a function of \LFIR. For comparisons, we include samples of local normal galaxies (including LIRGs; Brauher et al. 2008), local ULIRGs (Brauher et al. 2008; Farrah et al. 2013), $z>1$ SMGs (Maiolino et al. 2009; Ivison et al. 2010; De Breuck et al. 2011; Valtchanov et al. 2011; Coppin et al. 2012; Swinbank et al. 2012; Wagg et al. 2012; Riechers et al. 2013) and star-forming galaxies/composite systems (Stacey et al. 2010; Brisbin et al. 2015), and $z>1$ AGNs/QSOs (Pety et al. 2004; Maiolino et al. 2005; Stacey et al. 2010; Gallerani et al. 2012; Venemans et al. 2012; Wagg et al. 2012; Leipski et al. 2013; Wang et al. 2013; Willott et al. 2013). 

Figure~\ref{Figlineratio} illustrates one important feature --- the line-to-luminosity ratios are broadly dependent on \LFIR, and have very large variations at a given luminosity and redshift.  This large variation probably has to do with variations of ISM properties. For our sample, the \LOI/\LFIR\ and \LCII/\LFIR\  ratios are similar to that of local ULIRGs. This is not surprising because the \LFIR\ of our QSOs are only a factor of $(2-5)$ smaller than that of local ULIRGs and \LOI/\LFIR and \LCII/\LFIR ratios have intrinsic large scattering. However, we find that the line-to-luminosity ratios of our $z\sim0.15$ AGNs are significantly higher than that of AGNs/QSOs at $z$\,$\sim$\,1\,-\,6 (blue and purple symbols).  We emphasize this statement is not true for all high-$z$ AGNs in general, only true for these high-$z$ AGNs whose infrared luminosities are two orders of magnitude higher than our QSOs.  This is due to the fact that the published far-IR spectroscopy of high-$z$ QSOs is largely limited to extremely bright sources.

Finally,  Table~\ref{tableratio} and Figure~\ref{Figlineratio} may give an impression that \OI-to-luminosity ratios are higher for the AGN dominated QSOs (also with higher blackhole masses) than for the SB dominated ones.  The average ratio $\langle L_{\rm [O\,{\scriptsize \textsc{i}}]63\,\mu{\rm m}}/L_{\rm FIR}\rangle_{\rm AGN}$ is -2.49$\pm$0.14 for all detections, and -2.98$\pm$0.22 if we exclude W0752+19 (which is AGN dominated). The same average ratio is only -3.36$\pm$0.13 for the SB dominated QSOs.  It is possible such a trend between \LOI/\LFIR\ and AGN fraction does exist, however, it is very weak.  In addition, the average ratio $\langle L_{\rm [C\,{\scriptsize \textsc{ii}}]}/L_{\rm FIR}\rangle$ is $-2.90\pm0.20$ and $-3.12\pm0.35$ for the AGN- and SB-dominated systems respectively. These average ratios are statistically same for the two types of QSOs in our sample.

If the slightly enhanced \LOI/\LFIR\ ratios among the AGN dominated systems are real, this trend can not be true for all AGNs and star forming galaxies in general. It has to be limited to our selected QSOs which are at similar redshifts with similar \LIR\ and "effective" far-IR dust temperature (far-IR $f_{60}/f_{100}$ colors). A possible explanation is that AGN dominated systems, {\it i.e.} more massive blackholes, tend to have more emission at the mid-IR and much less at the far-IR. Table~\ref{tableratio} shows that the AGN dominated targets all have smaller \LFIR/\LIR\ ratios. On average, smaller \LFIR\ values lead to larger \LOI/\LFIR\  ratios. However, if this is the only explanation, it is hard to understand why \CII-to-luminosity ratios are similar among both types of QSOs in our sample. Another possible reason is that AGN dominated systems may have higher ionizing fluxes from stronger blackhole accretion activities, which in turn may produce more excited neutral O atoms, thus stronger \OIFIR\ emission, due to the fact that the J=1,0 levels of the O$^0$ ground states require 228K and 327K energy. This argument has been used by theoretical models (Meijerink \&\ Spaans 2005; Abel et al. 2009) to explain observed \CII/\OI\ ratios. However, the recent paper by Langer \&\ Pineda (2015) has argued differently, that X-rays in AGNs could produce more ions at higher ionization states, thus leading to low observed \CII\ fluxes among luminous ULIRGs and AGNs.  One of our QSOs, W1117+44, has an observed X-ray flux of $(2.24\pm0.06)\times10^{-12}$ erg s$^{-1}$ cm$^2$ (Bianchi et al. 2009). Comparing its X-ray luminosity of $1.24\times10^{44}$ erg s$^{-1}$ with the Langer \&\ Pindeda model (their Figure 14), we obtain a higher \LCII/\LFIR\ ratio (-2.69,-2.73,-2.79 for filling factors of 0.003,0.002,0.001, respectively; log scale) than our observed value ($<$$-3.26$; log scale). Thus, we conclude that it is not clear at all what the physical model is valid for explaining \CII\ and \OI\ emissions from  AGNs and galaxies. More observations in the \CII\ and \OI\ lines and X-ray for AGNs/QSOs would be helpful for understanding this contradictory.

Cross-checking the host galaxy morphologies shown in Figure~\ref{figsdssa}, \ref{figsdssb} against the line-to-luminosity ratios, we do not see any definitive correlation. This is probably due to the fact that the far-IR line ratios are more sensitive to the far-IR surface brightness, rather than directly the optical morphologies (Diaz-Santos et al. 2013). Our sources have typical QSO spectra with optical forbidden emission lines, such as \OIII\,5007\,\AA\ and \OIDOUBLE\, both coming from the Narrow Line Regions (NLRs) in these QSOs. The \OIII5007/\OI6300 line ratio is sensitive to ionization parameter.  The higher \OIII5007/\OI$\lambda$6300 ratios imply higher fractions of ionized O$^{++}$ atoms and the stronger ionization fields from the central AGNs. Here we use the observed ratio of $\log$(\OIII5007/H$\beta$)$-$$\log$(\OI$\lambda$6300/H$\alpha)$ as a proxy for \OIII5007/\OI$\lambda$6300, where the narrow component of H$\alpha$ and H$\beta$ and both the narrow and broad components of \OIII5007 and \OI$\lambda$6300 were used to obtain the ratio. In this way, the dust extinction is essentially taken into account by the ratio of H$\alpha$/H$\beta$.  Figure~\ref{Figionization} shows the derived \OIII5007/\OI$\lambda$6300 ratio versus \LOIFIR/\LFIR. From the figure we see a weak anti-correlation between the \LOI/\LFIR\ and \OIII5007/\OI$\lambda$6300 ratios, in the sense that the sources with smaller \OIII5007/\OI$\lambda$6300 ratios have higher fraction of \OI63\,\mum\ emission. This might indicate that a larger fraction of the ISM is in a neutral phase.

\begin{figure*}[pthb]
\centering
\includegraphics[width=0.85\textwidth,bb=28 20 440 316]{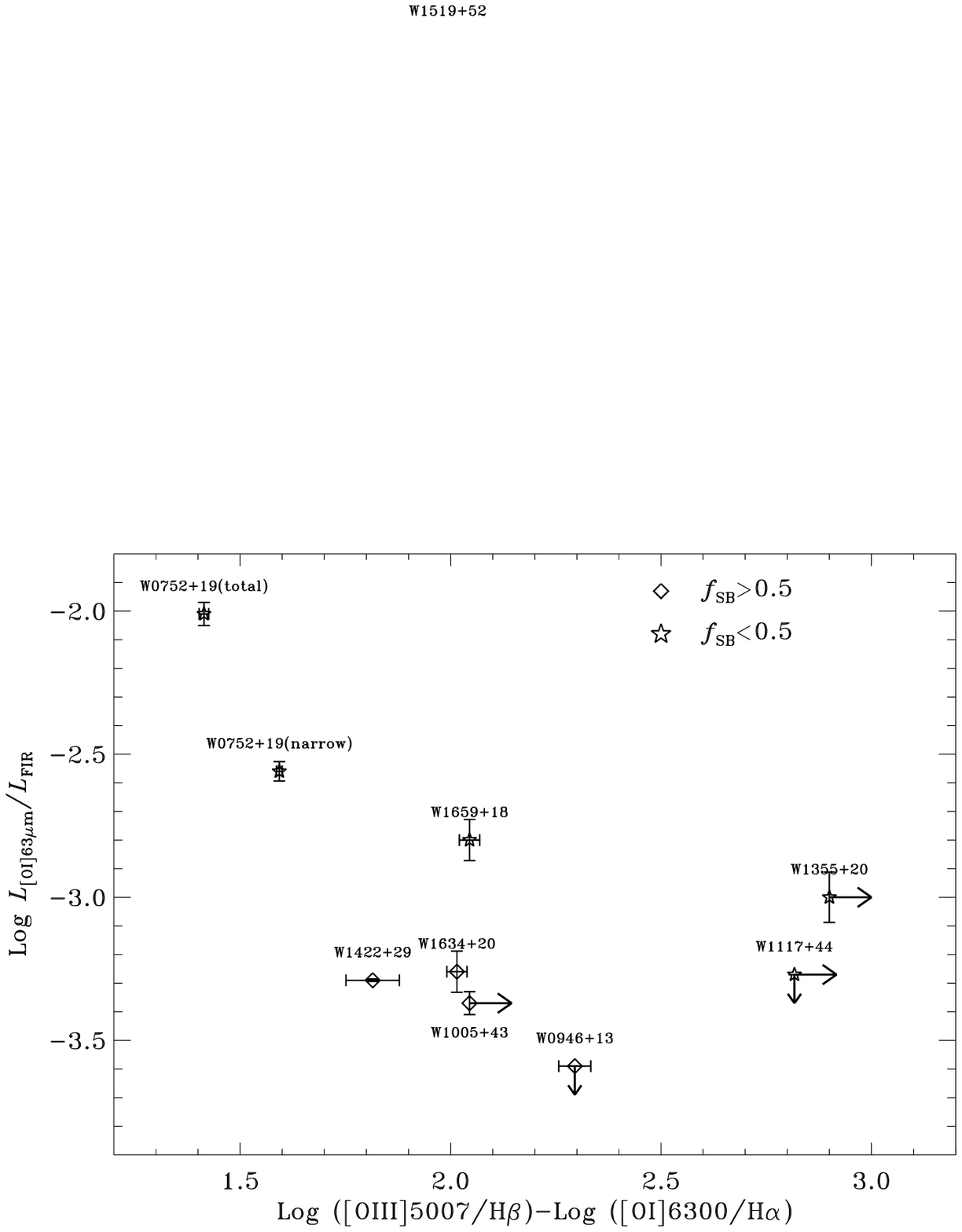}
\caption{\LOI/\LFIR\ versus \LOIII$_{\lambda5007}$/$L_{[\rm{O\,{\scriptsize \textsc {i}}}]\lambda 6300}$. The total and narrow components of W0752+19 are plotted separately in this figure.}
\label{Figionization}
\end{figure*}

\begin{figure*}[pthb]
\centering
\includegraphics[width=0.95\textwidth,bb=1 3 495 356]{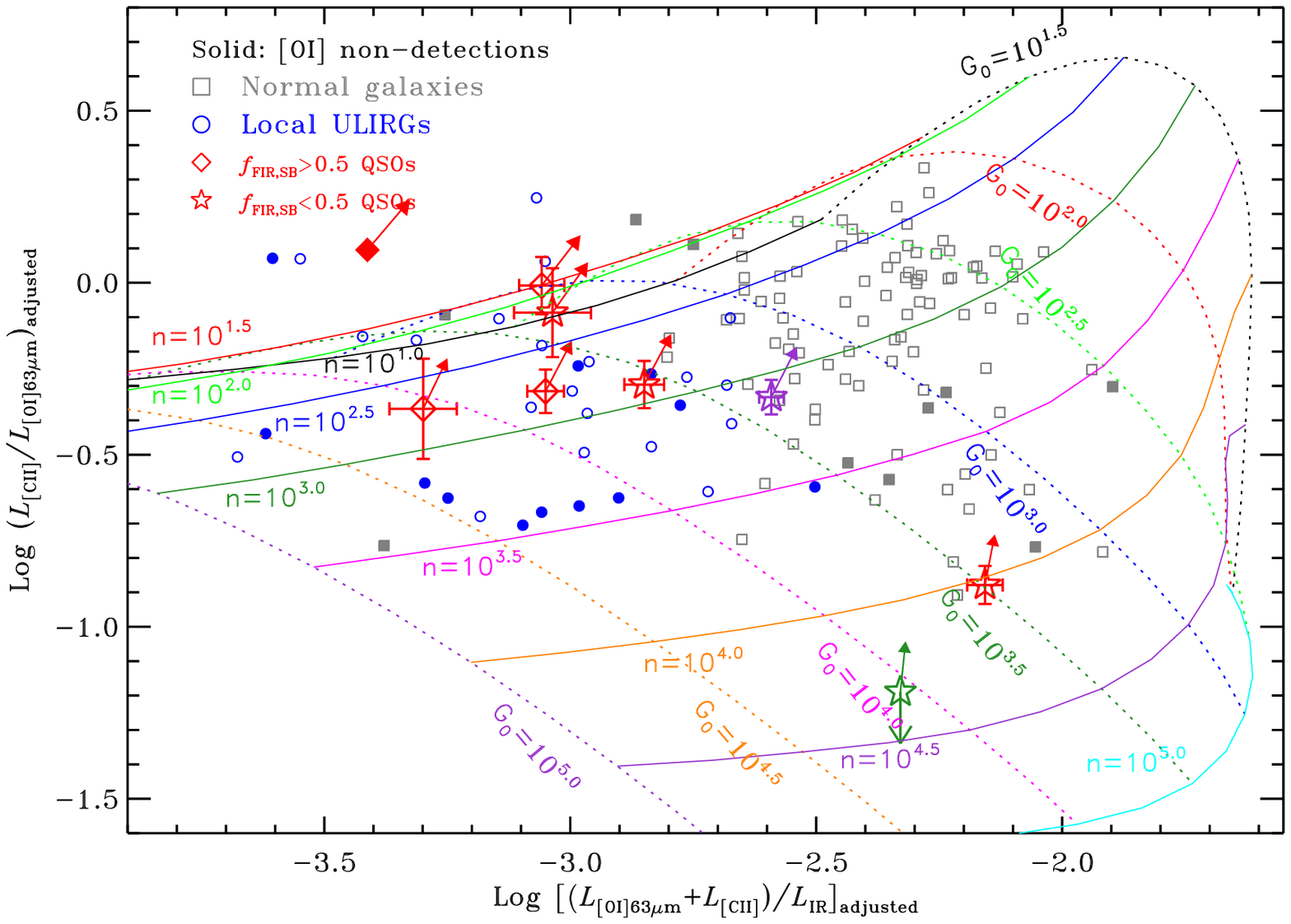}
\caption{Observed data points overlaid on the PDR model grid (Kaufman et al. 1999): dotted and solid lines are corresponding to constant strength of radiation field ($G_0$) and number density ($n$), respectively. The various geometric corrections have been applied to the observed \CII\ and \LIR\ values (see the text for details). Red symbols show QSOs in this work, with diamonds and stars representing $f_{\rm SB} > 0.5$ and $f_{\rm SB} < 0.5$ respectively. Solid arrows associated with our QSOs give the locations when the correction factor of 1.4 is not applied to the \CII\ flux. Solid symbols indicate that the \OIFIR\ line is undetected. The purple and green stars respectively show the narrow and broad components (but without splitting the IR flux) of W0752+19. In this figure we also plot local normal galaxies (including LIRGs; Brauher et al. 2008), and local ULIRGs (Brauher et al. 2008; Farrah et al. 2013) for comparisons.}
\label{Figpdrmodel}
\end{figure*}

One QSO, W0752+19, has exceptionally high \LOI/\LFIR\ ratio, about an order of magnitude higher than that of the other sources in our sample as well as that of local ULIRGs.  Its ratio is also higher than the average value for galaxies at similar \LFIR.  Like some rare cases at low-$z$, the \LOI/\LFIR\ ratio for W0752+19 is close to the high values observed among high-$z$ SMGs (Coppin et al 2012) and star forming galaxies (Brisbin et al. 2015) with \LFIR\ an order of magnitude higher. We discuss how and where such a strong emission is produced in W0752+19 in the following sections.

\subsubsection{Comparison to PDR Models}

In this section, we compare our observations with standard PDR models published in the literature. Strictly speaking, it is not appropriate to apply standard PDR models to the observations of IR bright QSOs because AGNs make contributions to both \OI\ and \CII\ emission as well as IR continuum luminosity.   Although we have done SED decomposition and can measure \LIR\ (SB) and \LIR\ (AGN) separately, we do not have sufficient information to do similar analysis on \OI\ and \CII. This underscores the urgent need to build suitable models for interpreting far-IR fine structure lines from AGNs/QSOs.  With these limitations in mind, we apply Kaufman et al. (1999) PDR models to our data. 

\OI-to-\CII\ ratios can be used to constrain ISM density and the strength of FUV radiation field. The \OIFIR\ has a much higher critical density for collisional excitation than that for \CII158\,\mum.  At the limit of very high density ($n \gg n_{\rm cr,[O\,{\scriptsize \textsc{i}}]}=5\times10^5$\,cm$^{-3}$), oxygen cooling is dominant,  both species are in local thermodynamic equilibrium and the ratio of their cooling rates is independent of density. In the low density limit (diffuse ISM, $n \ll n_{\rm cr,[C\,{\scriptsize \textsc{ii}}]}$$\sim$3000\,cm$^{-3}$), both cooling rates scale quadratically with density, thus the line ratio is independent of $n$. Only in the intermediate density regime, $n_{\rm cr,[C\,{\scriptsize \textsc{ii}}]}<n<n_{\rm cr,[O\,{\scriptsize \textsc{i}}]}$, the \OI-to-\CII\ ratio depends on density. 

One critical feature of Kaufman et al. (1999) PDR model is that it is built for a 2-dimensional slab with infinite length. However,  actual observations of distant extragalactic sources usually detect emissions from multiple, 3-dimensional PDR regions within a single beam.  Such a geometric difference implies the necessary corrections to \LOI, \LCII\ and \LIR.  To calculate the correct (\LOI$+$\LCII)/\LIR\ ratios for the model comparison, we use the following equation: (\LOI$+$\LCII/1.4)/(\LIR/2.0).  Here the observed \LIR\ is reduced by a factor of 2. This is because the actual observations detect infrared continuum emission from the front and back side of the clouds, especially when they are illuminated from all sides.  In contrast, the 2-D slab models only take into account the partial emission. The correction factor of 2 is suggested by Kaufman et al. (1999). 

The \OIFIR\ emission is optically thick, thus, the observed fluxes come from the front side of the clouds. The actual observed value is very similar to the model geometry of 2-D slab case. Thus no corrections are applied to the observed \OI\ fluxes.
In addition, the observed \CII\ emission is also corrected for geometric effect. As discussed in Kauffman et al. (1999) paper, \CII\ is thought to be marginally optically thick, and its optical depth $\tau$ at the line center is close to $\sim$1 (Luhman et al. 2003). Therefore, the observed \CII\ fluxes include emission mostly from the front side, and some from back side of the clouds. The observed values are (1.0+$e^{-\tau}$) times of that from a 2-D model, {\it i.e.} the observed fluxes divided by 1.4 when compared with 2-D models. However, this method is not always adopted in the published literature ({\it e.g.} Malhotra et al. 2001). Partly this is due to the complex nature of the problem where many factors such as optical depths, filling factors etc can easily affect the observed fluxes, whereas PDR models are still fairly simple. In later section, we also compare our observations with PDR models without correcting observed \CII\ fluxes.
 
Finally, one more correction applied to the observed \CII\ emission is to remove the contribution from extended, ionized gas. For this correction, we have utilized the relation between the ratio (\CII/\NIIIO) of the \CII\ to \NII\ 205 \mum\ emission and \fircolor\ (Lu et al. 2015), and adopted solar gas phase abundances of ${\rm C/H}=1.4\times10^{-4}$ and ${\rm N/H}=7.9\times10^{-5}$ (Savage \& Sembach 1996). Because the similar ionization potentials and the same critical densities (for collision partner of electron) for \CII\ and \NIIIO\ lines, the \CII/\NIIIO\ ratio, which is independent of the density, and mildly dependent on the ionization field intensity, can estimate the fraction of the \CII\ emission from ionized gas, based only on an assumed gas phase C/N abundance ratio (e.g. Oberst et al. 2006; Stacey et al. 2010). Here we adopt $\log\,[{\rm N}\,{\scriptsize \textsc{ii}}]205\,\mu{\rm m}/[{\rm C}\,{\scriptsize \textsc{ii}}] = -0.65 f_{60}/f_{100}-0.66$, which was obtained based on the observations of local (U)LIRGs (Lu et al. 2015). For our sample QSOs, the average \CII\ fraction from ionized medium is $\sim$0.1. Finally, we want to caution readers that these correction factors are uncertain and that non-PDR component (e.g., dusty ionized regions) might also contribute to the FIR continuum, especially for high $G_0$ sources (e.g. Luhman et al. 2003; Abel et al. 2009). 

Figure~\ref{Figpdrmodel} shows the comparison between the data and the calculated grid of density $n_0$ and FUV radiation field $G_0$ using the Kaufman et al. (1999) models. By assuming a local Galactic metallicity, these models were computed for a grid of density $n$ and FUV radiation field $G_0$ parameters with $n$=$10^1-10^7$\,cm$^{-3}$ and $G_0$=$10^{-0.5}-10^{6.5}$ in units of the local interstellar value, 1.6$\times$$10^{-3}$\,erg\,cm$^{-2}$\,s$^{-1}$ (Habing 1968). In Table \ref{table:pdr} we listed the derived values of $n$ and $G_0$, using the web-based tool PDR Toolbox{\footnotetext{\url{http://dustem.astro.umd.edu/pdrt/}}} (Kaufman et al. 2006; Pound \& Wolfire 2008). The uncertainties were obtained using Monte Carlo simulations by assuming a Gaussian error. In the table we also gave the results for which the AGN contribution to \LIR\ was removed.

\begin{deluxetable*}{lcccc}[pth]
\tablecaption{Results from PDR modeling\label{table:pdr}}
\tablewidth{0pt}
\tabletypesize{\scriptsize}
\tablehead{
\colhead{\multirow{2}{*}{Galaxy}}&\colhead{$\log\,n$}&\colhead{$\log\,G_0$}&\colhead{$\log\,n$}&\colhead{$\log\,G_0$}\\
\colhead{}&\colhead{(cm$^{-3}$)}&\colhead{($1.6\times10^{-3}$\,erg\,cm$^{-2}$\,s$^{-1}$)}&\colhead{(cm$^{-3}$)}&\colhead{($1.6\times10^{-3}$\,erg\,cm$^{-2}$\,s$^{-1}$)}\\
\colhead{(1)}&\colhead{(2)}&\colhead{(3)}&\colhead{(4)}&\colhead{(5)}
}
\startdata
W0752+19 (narrow)    &$3.25\pm0.09$ &$3.50\pm0.08$ &$3.50\pm0.29$ &$2.75\pm0.17$\\
W0752+19 (broad)   &$>$4.25     &$\sim$4.0 &     $>$5.25   & $\sim$1.0\\
W0752+19 (total) & $4.0\pm0.20$   & $ 3.50\pm0.17$   &$4.75\pm0.16$ & $0.75\pm11$\\
W0946+13 & $<$1.5  &  $<$3.25   &    $<$1.5  &   $<$2.75\\
W1005+43 &$3.00\pm0.17$ & $4.25\pm0.28$ &$3.00\pm0.28$&$4.00\pm0.28$\\
W1355+20{\tablenotemark{a, b}} & $1.25$ &$2.5$&\nodata&\nodata\\
W1422+29{\tablenotemark{c}} &$2.75\pm0.21$   &$3.75\pm0.17$& $2.75\pm0.21$ &$3.75\pm0.17$\\
W1634+20{\tablenotemark{a}} &$1.5$ &  $2.75\pm0.59$ &$2.5$ &$3.00\pm0.22$\\
W1659+18 &$2.75\pm0.22$ & $3.50\pm0.12$ &$3.25\pm0.35$&$2.75\pm0.20$\\
\enddata
\tablecomments{Column (1): source; Columns (2) and (3): density and FUV radiation field respectively; Column (4) and (5): density and FUV radiation field, respectively, with IR only from SB component.} 
\tablenotetext{a}{Values without errors mean that the uncertainty is very large since the observed data fall in regions where the PDR models show complicated behavior (see Figure \ref{Figpdrmodel}).}
\tablenotetext{b}{The AGN contribution to \LIR\ is 100\% and thus no fit to the SB component only.}
\tablenotetext{c}{The starburst contribution to \LIR\ is 100\%.}
\end{deluxetable*}

Excluding W0752+19, our sample have $n<$$\sim$$10^{3}$\,cm$^{-3}$ and the FUV radiation fields of $10^{3}\lesssim G_0 \lesssim 10^{4.2}$. Figure~\ref{Figpdrmodel} also compares our sample with local ULRIGs and normal star forming galaxies. It is clear that the ISM in our QSO sample has similar density and FUV radiation field ($G_0$) as in local Seyfert 1 ULIRGs (e.g. Farrah et al. 2013), but has much higher $G_0$ values than that of normal star forming galaxies. The higher $G_0$ values in QSOs and ULIRGs are expected due to the central powerful AGNs and/or compact nuclear starburst regions.} 

The exceptional source in our sample is W0752+19, which shows two components in its \OIFIR\ emission (Table \ref{tableratio}), and appears to be an ``outlier" in the \LOI/\LFIR-\LFIR\ relation (Figure \ref{Figlineratio}). The \OI\ broad velocity component does not have corresponding broad \CII\ component, detectable in our spectra. In Figure~\ref{Figpdrmodel}, we plot the both components separately, where the $3\sigma$ upper limit for the \CII\ broad component was adopted.  Please note that we do not split the IR flux since it is hard to do it in a fair way. Compared with the model grid, the narrow velocity component for W0752+19 has very similar properties as that of the rest of the sample with $n$$\sim$$10^{3.3}$ cm$^{-3}$ and $G_0$$\sim$$10^{3.5}$. However, the ISM of the broad velocity component has density of $>10^{4.3}$ cm$^{-3}$, much denser than any of local ULIRGs and normal galaxies, but reaching similar values observed among a few starburst galaxies at $z$$>$1.

In Figure \ref{Figpdrmodel} we also use solid arrows to indicate the locations ({\it i.e.} the arrow head) of our QSOs when the correction factor of 1.4 is not applied to the \CII\ flux. A source will move along both axes by 0.15 dex when the \CII\ emission dominates the cooling, and thus the estimated $n$ and $G_0$ will lower by $0.1-0.3$ dex.  It is important to note that in the regimes with very low or high $n$ values, small changes in line flux ratios could imply significant changes in the inferred density values from comparisons with models, as shown in Figure~\ref{Figpdrmodel}. However, one should be cautious because PDR models may not be very reliable in these regimes. Considering these $n$ and $G_0$ values are rough estimates, we believe that applying the correction factor of 1.4 to the \CII\ flux or not will not affect the main science conclusions in this paper.

\begin{figure*}[pthb]
\centering
\includegraphics[width=0.95\textwidth,bb=10 5 492 344]{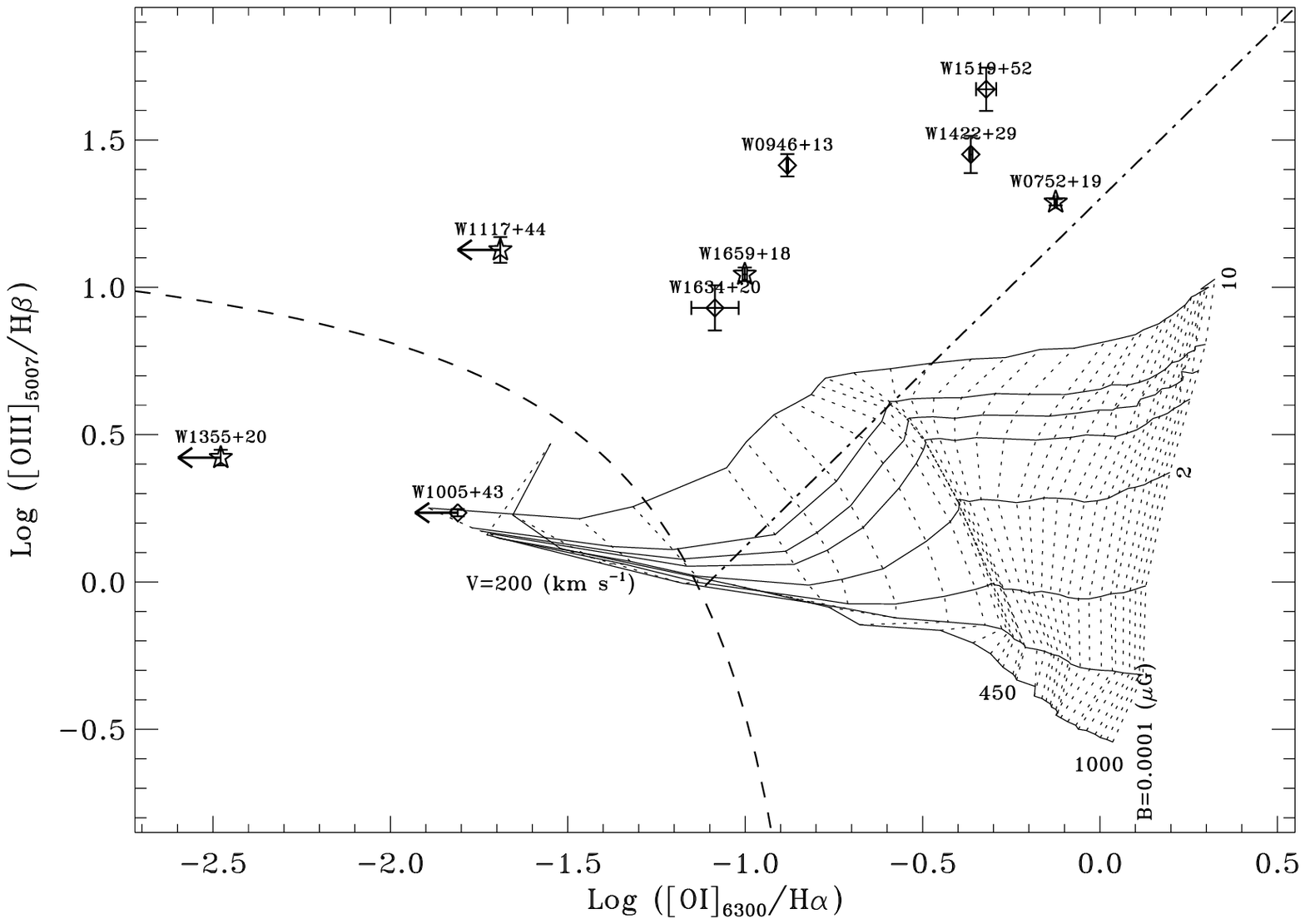}
\caption{Comparison of the observed ratios to the fast shock model grids with twice solar abundance and preschok density of ($n=1\,{\rm cm}^{-3}$) (Allen et al 2008) on the \OI$\lambda$6300/\Ha\ vs \OIII 5007/H$\beta$ diagram. Diamonds and stars respectively represent $f_{\rm SB}>0.5$ and $f_{\rm SB<0.5}$. Solid and dotted lines represent constant magnetic field ($B=10^{-4},\,0.5,\,1.0,\,2.0,\,3.23,\,4.0,\,5.0,\,{\rm and},\,10.0\, \mu {\rm G}$) and shock velocity ($v=200-1000\,{\rm km}^{-1}$), respectively.  The dashed line represents the Kewley et al. (2001) starburst/AGN classification line, while the dashed-dotted line shows the Seyfert-LINER division given by Kewley et al. (2006).}
\label{Figbpt}
\end{figure*}

\subsection{Origin of the FIR \OI\ and \CII\ Emissions}
\subsubsection{Is shock/turbulence an important heating source?}
The FIR fine structure lines \OIFIR\ and \CII\ are thought to be excited by collisions with atomic and molecular hydrogen in PDRs, where the gas are heated by the photo-electric effect, i.e., hot electrons are ejected from PAH molecules and small dust grains by absorbing FUV photos, and transfer their energy to the gas. \OIDOUBLE\ and \OIFIR\ emissions can also be produced or boosted by shocks (Hollenbach \& McKee 1989; Allen et al. 2008; Lesaffre et al. 2013) in the outflows/winds observed frequently in both nearby and distant QSOs (Greene et al. 2011; Sturm et al. 2011; Nesvadba et al. 2011; Cano-D\'{i}az et al. 2012; Harrison et al. 2012).

Specifically for the objects in our sample, the question is whether any observed FIR fine-structure emission could have different heating sources, particular the ones with broad velocity components. Could the dissipation of kinematic energy released through a cascade of turbulent motions from large to small scales be a significant heating source?  

To this end, we compared the observed \OI$\lambda$6300/\Ha\ ratios for the narrow component (i.e. ${\rm FWHM}<1200\,{\rm km s}^{-1}$) to the radiative shock models presented in Allen et al. (2008), as shown in Figure~\ref{Figbpt}. We have chosen the model grids which have twice solar abundance and pre-shock density of $n=1$ cm$^{-3}$, as discussed in Allen et al. (2008), because the photoionization models with this metallicity can well reproduce the observed narrow-line ratios in AGNs (Groves et al. 2004; 2006), and that AGNs are typically hosted in galaxies more massive than 10$^{10}$ $M_\odot$ and hence are likely to contain high-metallicity gas (e.g., Tremonti et al. 2004). From Figure \ref{Figbpt} we can see that shock models fall much too short to match the observed \OIDOUBLE\ ratios and the optical \OIDOUBLE\ emissions are likely powered by FUV photons from AGNs rather than shocks. 

In addition, observations and shock model calculations (Burton et al. 1990; Appleton et al. 2013; Lesaffre et al. 2013) have shown that \OI/FIR and \CII/FIR ratios for shocked gas tend to be elevated, about an order of magnitude higher than those produced by photoelectric heating in PDRs. As shown in \S3.3, the observed \OI/FIR (\CII/FIR) ratios for our QSOs do not show any significant enhancement comparing to normal/starburst/AGN galaxies, which further confirms that the gas emissions, both in the optical and FIR, should be heated by AGNs and/or starburst activities of the host galaxies.

\subsubsection{Where do FIR fine-structure lines come from?}
The basic result from our study is the detections of strong \OIFIR\ and \CII\ line emission from $z$$\sim$0.15 IR bright QSOs.  Using the far-IR line-to-luminosity ratios and the standard PDR models, we infer the densities of the warm neutral gas clouds responsible for the observed line emission are in the range of $10 - 10^3$\,cm$^{-3}$, with one exceptional case of W0752+19, whose broad \OIFIR\ component implies a density $>10^{4.3}$\,cm$^{-3}$.  

Sources with far-IR \OIFIR\ line widths as broad as 720\,km/s are rare. In the sample of local ULIRGs, Farrah et al. (2013) have found that about 30\%\ have broad components with velocity widths of $\sim$500\,km\,s$^{-1}$, none reached the level of 700\,km/s.  So where does this broad velocity component in W0752+19 come from?  There are three possibilities. The first is Narrow Line Region (NLR) with a size of a few hundred parsecs, photo-ionized by radiation field of an accreting supermassive blackhole.  The other two possibilities are star forming regions (PDRs) in the host galaxy and dense gas outflows. 

In the standard AGN/QSO model, the forbidden emission lines are collisionally excited and believed to come from NLRs with densities of $10^3-10^5$\,cm$^{-3}$ (see a review by Osterbrock 1991).  Most of our far-IR \OI\ and \CII\ detections indicate smaller densities, which suggest these narrow lines maybe mostly come from the PDRs in the host galaxies. With the same density argument, it would be consistent if the broad \OIFIR\ component in W0752+19 comes from the warm and dense ISM within the NLR. In addition, we find that the line centroids of broad far-IR \OI\ and broad optical \OI$\lambda$6300\,\AA\ are coinciding, with the negligible offset of 63\,km\,s$^{-1}$. This lends an additional support to the NLR origin for the broad \OIFIR\ emission in W0752+19. However, we note that the densities derived here are based on PDR models, which could be significantly different from those derived from a NLR model/geometry. Therefore, we can not totally rule out the NLR origin of these FIR fine-structure lines.

Gas outflows could be an alternative explanation since outflows are frequently observed in QSOs (e.g Fischer et al. 2010; Aalto et al. 2012; Cicone et al. 2014; Tadhunter et al. 2014). The broad and asymmetric (e.g extended blue wing) line profile seems to support this interpretation. The none-detection of the broad \CII\ emission line in W0752+19 means that the gas density of this outflow gas is quite high, $\ge10^4$\,cm$^{-3}$ (see \S3.3.2). At these high density regimes, \CII\ emission is suppressed due to collisional de-excitation. Such dense gas outflow is rare, and has been reported in Mrk\,231 (Aalto et al. 2012).  However, one potential difficulty in the outflow scenario is that we do not detect optical \OI$\lambda$6300\,\AA\ emission with a velocity width similar to the broad \OIFIR\ component in W0752+19.

\subsection{Implied Dynamical Masses}
There is another indirect and crude way to investigate the physical regime where the FIR \OI\ (\CII) emission may arise, {\it i.e.} to compare the dynamical mass estimates of the dense neutral gas ($M_{\rm dyn}$) for our sample with those having high spatial resolution observations, such as the high-$z$ QSOs in Wang et al. (2013), in which the \CII\ emissions were marginally resolved. Our QSOs have similar blackhole masses ($M_{\rm BH}$) as those in Wang et al. 2013. This comparison may shed some light on the properties of neutral ISM  in low and high-$z$ QSOs.

Many studies of AGNs/QSOs have proposed that NLRs could contain multi-components --- an inner biconical NLRs over a size of a few 100\,pc, and a much larger Extended Narrow Line Region (ENLR), which are gravity-dominated and could have organized velocity structures, such as outflows and rotations (Osterbrock 1991; Veilleux 1991). For a rotating gas disk with a radius $R$, the dynamical mass can be calculated as $M_{\rm dyn}$=$R*v_{\rm cir}^2/G$, where $v_{\rm cir}$ is the maximum circular velocity and $G$ the gravitational constant.  Because almost all of the detected \OIFIR\ and \CII\ lines are resolved, as shown in Table~\ref{tableline}, it is possible for us to construct a toy model, where we assume the neutral gas producing far-IR \OI\ and \CII\ emission is in a rotating disk with a radius of the size of ENLR, and we also approximate $v_{\rm cir}$$\sim$$\rm FWHM$(line)/$\sin i$, with $i$ as the disk orientation angle. We chose the larger value of the \OIFIR\ and \CII\ line widths for $v_{\rm cir}$.  The ENLR sizes can be estimated using the empirical $L_{5100\mathring{\mathrm{A}}}$-size relations, $\log(R_{\rm ENLR}[{\rm pc}]) = 0.46\log(L_{5100\mathring{\mathrm{A}}} {\rm [erg s^{-1}]}) -16.95$ (Bentz et al. 2009; Husemann et al. 2014). The $L_{5100\mathring{\mathrm{A}}}$ and black hole mass measurements are taken from Shen et al. (2011). The resulting ENLR radii are $1.4-5.1$\,kpc, which are comparable to sizes of host galaxies. This toy model yields dynamical masses $M_{\rm dyn}\sin^2 i$ of $(2.5\,-\,40)$\,$\times10^{10}M_\odot$.

Is this toy model consistent with the gas densities derived from the \OIFIR/\CII\ ratios? As we showed before, the gas densities are mostly in the range of $10-10^3$\,cm$^{-3}$. If we take the gas density of $10^3$\,cm$^{-3}$, the disk scale height $H$ of 200\,pc, the gas disk with the radii of ENLRs, and the volume filling factor of 0.1, we obtain the total gas masses $M_{\rm H+H_2} = \pi R^2 H\mu_{\rm H}m_{\rm H} n_{\rm H} = 4\times10^9 - 5\times10^{10}M_\odot$, with $\mu_{\rm H}=1.3$ as mean atomic weight for neutral gas. These numbers are somewhat smaller, but consistent with the dynamical estimates.

How are our dynamical mass estimates compared with the measurements for high-$z$ QSOs? Wang et al. (2013) recently published a sample of $z$$\sim$5 QSOs with the ALMA \CII\ maps which spatially resolved \CII\ emitting gas, yielding the sizes of $2.6-5.3$\,kpc. Their derived $M_{\rm dyn}\sin^2i$ are in the range of 1.6\,-\,15$\times10^{10}M_\odot$ using the same definition of $v_{\rm cir}$ as ours.  This is similar to the dynamical gas masses derived for our QSOs.  At the face value, this similarity implies that the masses of dense neutral ISM are not changing significantly in QSOs at $z$\,$\sim0.15$ and $z$\,$\sim5$.  This inference seems to be in contradiction with other published ISM studies, suggesting that high-$z$ sources have higher gas fractions relative to their stellar masses (Tacconi et al. 2013; Genzel et al. 2015).  This contradiction implies that our assumption of radii $R$ in our calculation may be too large, {\it i.e.}, the sizes of the far-IR line emitting regions in low-$z$ QSOs are likely smaller than that of ENLRs, $1.4-5.1$\,kpc.   Future high spatial observations are required to confirm our speculations.

\section{Summary}
We present \herschel\ PACS spectroscopic observations of a sample of nine QSOs, selected to be at $z$$\sim$0.1\,-\,0.2 from {\it SDSS} spectra and have 22\mum\ fluxes greater than 1\,mJy.  The full IR SED fitting reveals that our QSOs are mostly starburst and AGN composite systems, with  \LFIR\ around several times $10^{11}L_\odot$, a factor of a few less than that of local ULIRGs and over an order of magnitude less luminous than that of $z$$>$1 QSOs with available FIR spectroscopy.  Our PACS/\herschel\ observations detect strong \OIFIR\ and \CII158\,\mum\ emission, and discover one object, W0752+19, with exceptionally strong \OIFIR\ emission and broad component (720\,km/s).  We find that the observed \OIFIR/FIR and \CII/FIR ratios have similar distributions as that of local ULIRGs, but much higher than that of $z$$>$1 QSOs/AGNs with an order of magnitude higher FIR luminosities. Our observations also suggest that the AGN dominated systems in our sample seem to have higher \OIFIR/FIR ratios, which can be explained by the hotter SEDs produced by AGNs, and by the lower \LFIR/\LIR\ ratios in these systems.

Assuming the \OIFIR\ and \CII\ lines arise mainly in PDRs, we use theoretical models to constrain the hydrogen density, $n$, and FUV radiation field, $G_0$, in the FIR line emitting gas. We find $n \lesssim 10^{3.3}$ cm$^{-3}$ and $10^3\lesssim G_0 \lesssim 10^{4.2}$ for all systems except for the broad component emitting gas of W0752+19, which has $n \gtrsim 10^{4.3}$ cm$^{-3}$ and $G_0>10^4$. The former ranges of $n$ and $G_0$ are consistent with those found in local Seyfert 1 ULIRGs (Farrah et al. 2013). Our analysis of the optical \OI$\lambda$6300\,\AA\ emission suggests that shocks do not contribute significantly to the gas heating in our QSO sample.

\begin{acknowledgements}
We thank the anonymous referee for her/his careful reading of the manuscript and the constructive comments/suggestions, which have helped to improve the paper. We also thank Dr M. G. Wolfire for helpful discussion. Y.Z. is partially supported by the National Natural Science Foundation of China under grants No. 11390373 and 11420101002, and the CAS pilot-b project \#XDB09000000. This research has made use of the NASA/IPAC Extragalactic Database (NED), which is operated by the Jet Propulsion Laboratory, California Institute of Technology, under contract with the National Aeronautics and Space Administration. This publication makes use of data products from the {\it Wide-field Infrared Survey Explorer}, which is a joint project of the University of California, Los Angeles, and the Jet Propulsion Laboratory/California Institute of Technology,  funded by the National  Aeronautics  and  Space  Administration.  This  paper also  utilized  the  publicly  available  SDSS  data  sets.  Funding for  the  SDSS  and  SDSS-II  has  been  provided  by  the  Alfred P.  Sloan  Foundation,  the  Participating  Institutions,  the  National Science Foundation, the US Department of Energy, theNational Aeronautics and Space Administration, the Japanese Monbukagakusho, the Max Planck Society, and the Higher Education Funding Council for England. The SDSS Web site is http://www.sdss.org/. The SDSS is managed by the Astrophysical Research Consortium for the Participating Institutions. The Participating Institutions are the American Museum of Natural History, Astrophysical Institute Potsdam, University of Basel, University  of  Cambridge,  Case  Western  Reserve  University, University of Chicago, Drexel University, Fermilab, the Institute for Advanced Study, the Japan Participation Group, Johns Hopkins University, the Joint Institute for Nuclear Astrophysics, the  Kavli  Institute  for  Particle  Astrophysics  and  Cosmology, the Korean Scientist Group, the Chinese Academy of Sciences (LAMOST), Los Alamos National Laboratory, the Max-Planck-Institute for Astronomy (MPIA), the Max-Planck-Institute for Astrophysics (MPA), New Mexico State University, Ohio State University, University of Pittsburgh, University of Portsmouth, Princeton University, the United States Naval Observatory, and the University of Washington.
\end{acknowledgements}


\end{document}